\documentclass[fleqn,usenatbib]{mnras}

\usepackage{newtxtext,newtxmath}
\usepackage{longtable}
\usepackage{verbatim, longtable}	
\usepackage[figuresright]{rotating}

\usepackage[T1]{fontenc}
\usepackage{CJKutf8}

\def\mgii{{Mg~\sc II}}

\DeclareRobustCommand{\VAN}[3]{#2}
\let\VANthebibliography\thebibliography
\def\thebibliography{\DeclareRobustCommand{\VAN}[3]{##3}\VANthebibliography}

\usepackage{graphicx}	
\usepackage{amsmath}	

\title[jets of BL Lacs]{Jet power extracted from ADAFs and the application to Fermi BL Lacertae objects}

\author[Yongyun Chen et al.]{Yongyun Chen\begin{CJK*}{UTF8}{gkai}(陈永云)\end{CJK*}\thanks{E-mail: ynkmcyy@yeah.net}$^{1}$,
Qiusheng Gu\begin{CJK*}{UTF8}{gkai}(顾秋生)\end{CJK*}\thanks{E-mail: qsgu@nju.edu.cn}$^{2}$,
Junhui Fan\begin{CJK*}{UTF8}{gkai}(樊军辉)\end{CJK*}$^{3}$,
Xiaoling Yu \begin{CJK*}{UTF8}{gkai}(俞效龄)\end{CJK*}$^{1}$,
\and Nan Ding\begin{CJK*}{UTF8}{gkai}(丁楠)\end{CJK*}$^{4}$,
Xiaotong Guo \begin{CJK*}{UTF8}{gkai}(郭晓通)\end{CJK*}$^{5}$,
Dingrong Xiong\begin{CJK*}{UTF8}{gkai}(熊定荣)\end{CJK*}$^{6}$
\\
$^{1}$College of Physics and Electronic Engineering, Qujing Normal
University, Qujing 655011, P.R. China\\
$^{2}$School of Astronomy and Space Science, Nanjing University, Nanjing 210093, P. R. China\\
$^{3}$Center for Astrophysics,Guang zhou University,Guang zhou510006, China\\
$^{4}$School of Physical Science and Technology, Kunming University 650214, P. R. China\\
$^{5}$School of mathematics and physics, Anqing Normal University 246011, P. R. China\\
$^{6}$Yunnan Observatories, Chinese Academy of Sciences, Kunming 650011, China\\
}

\date{Accepted XXX. Received YYY; in original form ZZZ}

\pubyear{2023}

\begin{document}
\label{firstpage}
\pagerange{\pageref{firstpage}--\pageref{lastpage}}
\maketitle

\begin{abstract}
We calculate the jet power of the Blandford-Znajek (BZ) model and the hybrid model based on the self-similar solution of advection-dominated accretion flows (ADAFs). We study the formation mechanism of the jets of BL Lacs with known redshifts detected by the Fermi satellite after 10 yr of data (4FGL-DR2). The kinetic power of the jets of Fermi BL Lacs is estimated through radio luminosity. The main results are as follows. (1) We find that the jet kinetic power of about 72\% intermediate peak frequency BL Lacs (IBL) and 94\% high-frequency peak BL Lacs (HBL) can be explained by the hybrid jet model based on ADAFs surrounding Kerr black holes. However, the jet kinetic power of about 74\% LBL cannot be explained by the BZ jet model or the hybrid model. (2) The LBL has a higher accretion rate than IBL and HBL. About 14\% IBL and 62\% HBL have pure optically thin ADAFs. However, 7\% LBL may have a hybrid structure consisting of an standard thin disk (SS) plus optically thin ADAFs. (3) After excluding the redshift dependence, there is a weak correlation between the jet kinetic power and the accretion disk luminosity for Fermi BL Lacs. (4) There is a significant correlation between inverse Compton luminosity and synchrotron luminosity for Fermi BL Lacs. The slope of the relation between inverse Compton luminosity and synchrotron luminosity for Fermi BL Lacs is consistent with the synchrotron self-Compton (SSC) process. The result may suggest that the high-energy components of Fermi BL Lacs are dominated by the SSC process. 
\end{abstract}




\section{Introduction}
BL Lacertae objects (BL Lacs) are a subclass of the blazar, an extreme type of active galactic nuclei (AGN), whose jets point to observer \citep[e.g.,][]{Blandford1978}. The difference between BL Lacs and their sibling flat-spectrum radio quasars (FSRQs) is that the spectra of BL Lacs lack emission lines with an equivalent width less than 5\AA~ \citep[e.g.,][]{Urry1995}. BL Lacs usually have only weak or non-existent emission lines. The lack of strong emission lines may be attributed to an ineffective accretion process, which does not produce sufficient energy to photoionize the broad line region (BLR) clouds \citep[e.g.,][]{Ghisellini2011, Sbarrato2014}. \cite{Ghisellini2011} found that the BL Lacs have a low ratio of the luminosity of the broad-line region (BLR) to the Eddington luminosity, $L_{\rm BLR}/L_{\rm Edd}\leq5\times10^{-4}$. \cite{Sbarrato2014} also found that the BL Lacs have $L_{\rm BLR}/L_{\rm Edd}<10^{-3}$. These results show that the BL Lacs have a low accretion rate. 

According to the frequency at which the synchrotron component of the spectral energy distribution (SED) peaks, BL Lacs can be divided into different subcategories, namely, low-frequency peak BL Lacs (LBL), intermediate peak frequency BL Lacs (IBL) and high-frequency peak BL Lacs (HBL; \cite{Padovani1996}), and set the boundaries as $\log\nu_{\rm p}<14$ Hz for LBL,  14 Hz$<\log\nu_{\rm p}<15$ Hz for IBL, and $\log\nu_{\rm p}>15$ Hz for HBL \citep[e.g.,][]{Abdo2010}. Since July 2008, the large area telescope on the Fermi Gamma-Ray Space Telescope (LAT, \cite{Atwood2009}) has scanned the entire gamma-ray sky about every three hours. 
Many blazars have been detected to have high-energy gamma-ray emissions, and the research of blazars has entered a new era. The LAT AGN catalog shows that the BL Lacs are the largest group of $\gamma$-ray sources \citep{Abdollahi2020, Ajello2020}. The most recent LAT AGN catalog contains 1207 BL Lacs (4FGL-DR2, \cite{Abdollahi2020, Ajello2020, Foschini2022}), which makes it possible to study the physical properties of a $\gamma$-ray selected sample of BL Lacs. \cite{Li2010} studied the relationship between optical-to-X-ray ($\alpha_{\rm ox}$) and X-ray-to-$\gamma$-ray ($\alpha_{\rm x\gamma}$) composite spectral indices and found that FSRQs and LBL occupy the same region by using 54 Fermi blazars. They suggested that FSRQs and LBL have similar spectral properties.

By now, several questions have been raised about BL Lacs, for example, what is the formation mechanism of the jets of BL Lacs? There are three theories of jet formation. The first is the Blandford-Znajek (BZ) mechanism \citep{Blandford1977}, in which jets extract the rotational energy of black holes. The second is the  Blandford-Payne (BP) mechanism \citep{Blandford1982}, in which the jet extracts the rotational energy of the accretion disk. In the above two cases, it should be maintained by matter accreting on the black hole, which leads to an expected relationship between accretion disk luminosity and jet power \citep{Maraschi2003}. Many authors have confirmed this relationship \citep[e.g.,][]{Rawlings1991, Cao1999, Wang2004, Ghisellini2009, Gu2009, Ghisellini2010, Ghisellini2011, Sbarrato2012, Sbarrato2014, Xiong2014, Chen2015b, Paliya2017, Paliya2019, Xiao2022, Zhang2022}. The third is the hybrid model, that is, the combination of BZ and BP mechanisms \citep{Meier1999, Meier2001, Garofalo2010}. \cite{Garofalo2010} used a hybrid model to speculate the differences observed in AGN with relativistic jets. \cite{Cao2003} calculated the maximal jet power of BP mechanism and BZ mechanism for a standard thin accretion disk. Comparing with the jet kinetic power of 29 BL Lacs, they found that the BZ and BP mechanisms could not explain the jet kinetic power of BL Lacs. These results imply that the accretion disks in most BL Lacs should not be standard accretion disks. \cite{Deng2021} used the one-zone Leptonic jet model to get the jet kinetic power of two HBLs (Mrk 421 and Mrk 501). Compared with the maximum jet power of the BZ mechanism and the BP mechanism for a thin disk, they found that the BZ mechanism may explain the jet kinetic power of Mrk 421, while the jet kinetic power of Mrk 501 may be explained by the BP mechanism or the BZ mechanism. \cite{Xiao2022} found a significant correlation between jet power and normalized disk luminosity ($L_{\rm disk}/L_{\rm Edd}$) for 16 BL Lacs. They suggested that the jet powers of 16 BL Lacs are powered by BZ mechanism. Although there are some studies on the jet mechanism of BL Lacs, however, no author has used a large sample of BL Lacs to study their jet mechanism in the case of ADAFs.    

The SED contains two emission components, namely the synchrotron component and inverse Compton (IC) component \citep[e.g.,][]{Ghisellini1997, Massaro2004, Massaro2006}. In the Leptonic model, it is generally believed that the low energy peak is generated by the synchrotron emission of relativistic electrons from the jet and the high energy peak is generated by IC scattering \citep[e.g.,][]{Massaro2004, Massaro2006, Meyer2012}. However, there are disagreements regarding the origin of soft photons scattered by IC. (1) They come from synchrotron emission, called synchrotron self Compton (SSC) process \citep[e.g.,][]{Rees1967, Jones1974, Marscher1985, Maraschi1992, Sikora1994, Bloom1996}. (2) They come from the outside of the jet, called the external Compton (EC) process. There are three possible sources of EC soft photons: accretion disk photons entering the jets directly \citep{Dermer1992, Dermer1993}; broad-line region (BLR) photons entering the jets \citep{Sikora1994, Dermer1997}; and infrared radiation photons from the dust torus entering the jet \citep{Blazejowski2000, Arbeiter2002}. \cite{Ghisellini1996} obtains two relationships between synchrotron luminosity and IC luminosity, which determines whether the IC component is dominated by the EC process or the SSC process ($L_{\rm EC}\sim L_{\rm syn}^{1.5}$, $L_{\rm SSC}\sim L_{\rm syn}^{1.0}$).        

 In this work, we study the physical properties of the jets of BL Lacs. Section 2 describes the sample. Section 3 is the jet model. In Section 4, we describe the results and discussion. Section 5 is the conclusion. Throughout this article, a standard concordance cosmology was assumed ($H_{0}=70$ km s$^{-1}$Mpc$^{-1}$, $\Omega_{\rm M}=1-\Omega_{\rm A}=0.27$).

\section{The Fermi BL Lacs sample}
We try to select a large sample of Fermi BL Lacs with reliable redshift, black hole mass, and accretion disk luminosity. For this, we consider the sample of \cite{Paliya2021}, who used 1077 blazars detected with the Fermi Large Area Telescope (4FGL-DR2) to study the central engines of Fermi blazars. 

\subsection{The black hole mass}
For the case of BL Lac objects that lack strong broad emission lines, \cite{Paliya2021} obtained the black hole mass ($M_{\rm BH}$) through the following two methods.

First, the black hole mass is calculated by using the stellar velocity dispersion ($\sigma_{*}$). The formula is as follows \citep{Gultekin2009},  

\begin{equation}
	\log \left(\frac{M_{\rm BH}}{M_{\odot}}\right) = (8.12\pm0.08) + (4.24\pm0.41)\times\log\left(\frac{\sigma_{*}}{200 \rm km~s^{-1}}\right).
\end{equation}

Second, \cite{Paliya2021} used the bulge luminosity to estimate the black hole mass. The formula is as follows \citep{Graham2007},

\begin{equation}
	\log \left(\frac{M_{\rm BH}}{M_{\odot}}\right) = \left\{
	\begin{array}{cc}
		(-0.38\pm0.06)(M_{\rm R}+21)+(8.11\pm0.11),\\
		(-0.38\pm0.06)(M_{\rm K}+24)+(8.26\pm0.11).
	\end{array}
	\right.
\end{equation}
where $M_{\rm R}$ and $M_{\rm K}$ are the absolute magnitudes of the host
galaxy bulge in the $R$ and $K$ bands, respectively.

We also note that the black hole mass of our sample is obtained by different methods. The uncertainty of the black hole mass estimated by stellar velocity dispersion is small, $\leq$ 0.25 dex.  The uncertainty of the black hole mass estimated by the bulge luminosity is 0.6 dex.       

\subsection{The accretion disk luminosity}
Due to the lack of detection of  emission lines, it is impossible to use emission lines to infer the BLR luminosity. Therefore, \cite{Paliya2021} derived a 3$\sigma$ upper limit in the  H$\beta$ (or \mgii, depending on the source redshift and wavelength coverage) line luminosity. \cite{Paliya2021} calculated the BLR luminosity by scaling several strong emission lines to the quasar template spectrum of \cite{Francis1991} and \cite{Celotti1997}, using Ly$\alpha$ as a reference and giving the total BLR fraction $<L_{\rm BLR}> = 555.77$. The BLR luminosity is estimated through the following formula,

\begin{equation}
L_{\rm BLR} = L_{\rm line}\times \frac{<L_{\rm BLR}>}{L_{\rm ref.frac.}},
\end{equation} 
where $L_{\rm line}$ is the emission-line luminosity and $L_{\rm ref.frac.}$ is the ratio: 22 and 34 for H$\beta$/Ly$\alpha$ and \mgii/Ly$\alpha$ \citep{Francis1991, Celotti1997}, respectively. When more than one line luminosity measurements were available, \cite{Paliya2021} took their geometric mean to derive the average $L_{\rm BLR}$. The accretion disk luminosity is estimated by using $L_{\rm disk}= 10L_{\rm BLR}$ \citep[e.g.,][]{Baldwin1978}, with an average uncertainty of a factor 2 \citep{Calderone2013, Ghisellini2014}.  

We carefully examined the samples of \cite{Paliya2021} and compared them with the classification of the sources of \cite{Abdollahi2020} and \cite{Foschini2022}.  We only consider that these sources with 1.4 GHz radio flux comes from the The FIRST Survey Catalog \citep{Becker1995}: 14Dec17 Version\footnote{http://sundog.stsci.edu/first/catalogs/readme.html}. Finally, we get 276 Fermi BL Lacs (57 LBL, 43 IBL, and 176 HBL). The relevant data is listed in Table 1.     

\subsection{The jet kinetic power}
\cite{Komossa2018} used the following formula to estimate the jet kinetic power ($P_{\rm kin}$) in AGN \citep{Birzan2008}

\begin{equation}
\log P_{\rm kin} = 0.35(\pm0.07)\log P_{1.4} + 1.85(\pm0.10)
\end{equation}
where $P_{\rm kin}$ is in units of $10^{42}$ erg s$^{-1}$, and $P_{1.4}$ is the 1.4 GHz radio luminosity in units of $10^{40}$ erg s$^{-1}$, $P_{1.4}=4\pi d_{L}^{2}\nu S_{\nu}$. The scatter for this relation is $\sigma=0.85$ dex. We make a K-correction for the observed flux using $S_{\nu}=S_{\nu}^{obs}(1+z)^{\alpha-1}$, where $\alpha$ is the spectral index and $\nu$ is the frequencies, and $\alpha=0$ is adopted \citep{Abdo2010, Komossa2018}. The $S_{\nu}^{obs}$ is the observed flux.  The $d_{L}$ is the distance of luminosity, $d_{L}(z)=\frac{c}{H_{0}}(1+z)\int_{0}^{z}[\Omega_{\rm A}+\Omega_{\rm M}(1+z^{'})^{3}]^{-1/2}dz^{'}$ \citep{Venters2009}. Most galaxies have a central supermassive black hole, which may coevolve with the host galaxy, resulting in correlations between bulge luminosity, stellar velocity dispersion, and central black hole mass \citep[e.g.,][]{Kormendy1995, Magorrian1998}. Models show that these correlations are caused by galaxy merging and feedback from AGN \citep[e.g.,][]{Silk1998, Kauffmann2000}. Around the time of these discoveries, the Chandra X-ray observation found direct evidence of AGN feedback, when observations revealed cavities and shock fronts in X-ray emission gas around many massive galaxies \citep[e.g.,][]{Fabian2000, McNamara2000}. The X-ray cavity provides a direct measurement of the mechanical energy released by the AGN through the work done on the hot, gaseous halo around them \citep{McNamara2000}. This energy is expected to heat the gas and prevent it from cooling and forming stars \citep{Churazov2001}. Studies on X-ray cavities show that AGN feedback provides enough energy to regulate star formation and inhibit the cooling of the hot halos of galaxies and clusters \citep[e.g.,][]{Birzan2008, Komossa2018}. \cite{Komossa2018} believes that if the relationship of \cite{Cavagnolo2010} is used to calculate jet kinetic power, the high value of jet kinetic power is predicted up to an order of magnitude. At the same time, all of our samples have a 1.4 GHz radio flux, but there is almost no low-frequency radio flux (for example 150 MHz). We cannot use low-frequency radio flux to calculate jet kinetic power. Therefore, we follow the method of \cite{Komossa2018} and use equation (1) to calculate the jet kinetic power. We note that the total jet power ($P_{\rm jet}$) is given by the sum of two components i.e., the radiation power ($P_{\rm rad}$) and the kinetic power ($P_{\rm kin}$). The jet power of radiation is believed to be about 10\% of the jet kinetic power \citep{Nemmen2012, Ghisellini2014}, namely $P_{\rm kin}\sim 10P_{\rm rad}$. Since the total jet power is mainly dominated by the jet kinetic power, the value of the total jet power can be approximately equal to the jet kinetic power. 

\begin{table*}
	\begin{minipage}{150mm}
		\centering
		\caption{The sample of Fermi BL Lacs.}
		\begin{tabular}{@{}crcccccccccccccccrl@{}}
			\hline\hline
			4FGL name & RA & DEC & Type & Redshift & $\rm{\log M}$ & $\log L_{\rm disk}$ & $f_{\nu}$ & $\log P_{\rm kin}$ & $\log L_{\rm sy}$ & $\log L_{\rm ic}$ \\
			{(1)} & {(2)} & {(3)} & {(4)} & {(5)} & {(6)} & {(7)} & {(8)} & {(9)} & {(10)} & {(11)}\\
			\hline
J0003.2+2207	&	0.8058	&	22.1302	&	HBL	&	0.1	&	8.10 	&	42.74 	&	0.0087	&	43.84 	&	43.23 	&	42.55 	\\
J0006.3-0620	&	1.5992	&	-6.3493	&	LBL	&	0.347	&	8.92 	&	44.52 	&	2.051	&	45.12 	&	45.74 	&	44.79 	\\
J0013.9-1854	&	3.4804	&	-18.9118	&	HBL	&	0.095	&	9.65 	&	43.27 	&	0.0295	&	44.01 	&	44.02 	&	42.94 	\\
J0014.1+1910	&	3.5368	&	19.1713	&	LBL	&	0.477	&	7.47 	&	44.32 	&	0.154	&	44.86 	&	45.46 	&	45.61 	\\
J0014.2+0854	&	3.5695	&	8.9114	&	HBL	&	0.163	&	8.85 	&	43.37 	&	0.326	&	44.55 	&	43.75 	&	43.38 	\\
J0015.6+5551	&	3.9071	&	55.8636	&	HBL	&	0.217	&	9.68 	&	44.05 	&	0.0849	&	44.45 	&	44.67 	&	44.21 	\\
J0017.8+1455	&	4.4711	&	14.9228	&	IBL	&	0.303	&	8.27 	&	44.16 	&	0.0595	&	44.52 	&	44.84 	&	44.53 	\\
J0021.6-0855	&	5.4115	&	-8.9174	&	IBL	&	0.648	&	8.54 	&	44.63 	&	0.0472	&	44.82 	&	45.33 	&	45.43 	\\
J0022.0+0006	&	5.5154	&	0.1134	&	HBL	&	0.306	&	8.02 	&	43.79 	&	0.0042	&	44.13 	&	44.57 	&	44.06 	\\
J0032.4-2849	&	8.1076	&	-28.8224	&	LBL	&	0.324	&	8.47 	&	44.02 	&	0.161	&	44.70 	&	44.95 	&	45.05 	\\
J0040.4-2340	&	10.1012	&	-23.6704	&	IBL	&	0.213	&	8.68 	&	43.75 	&	0.0536	&	44.38 	&	44.00 	&	43.74 	\\
J0045.7+1217	&	11.4309	&	12.292	&	HBL	&	0.255	&	8.82 	&	44.18 	&	0.104	&	44.54 	&	44.78 	&	44.65 	\\
J0049.0+2252	&	12.252	&	22.8735	&	IBL	&	0.264	&	9.04 	&	43.63 	&	0.076	&	44.51 	&	44.09 	&	43.88 	\\
J0056.3-0935	&	14.0874	&	-9.5997	&	HBL	&	0.103	&	8.96 	&	43.22 	&	0.201	&	44.32 	&	43.92 	&	43.52 	\\
J0059.3-0152	&	14.8361	&	-1.8725	&	HBL	&	0.144	&	8.63 	&	43.52 	&	0.018	&	44.07 	&	44.15 	&	43.68 	\\
J0103.5+1526	&	15.8786	&	15.4348	&	IBL	&	0.246	&	9.02 	&	43.78 	&	0.226	&	44.65 	&	44.25 	&	43.94 	\\
J0103.8+1321	&	15.969	&	13.3536	&	HBL	&	0.49	&	9.69 	&	44.41 	&	0.0519	&	44.70 	&	45.33 	&	44.92 	\\
J0105.1+3929	&	16.2913	&	39.4963	&	LBL	&	0.44	&	8.17 	&	44.34 	&	0.0915	&	44.74 	&	45.76 	&	46.05 	\\
J0111.4+0534	&	17.8573	&	5.5761	&	HBL	&	0.347	&	8.47 	&	44.06 	&	0.0165	&	44.38 	&	44.82 	&	43.97 	\\
J0115.8+2519	&	18.9539	&	25.3324	&	HBL	&	0.376	&	9.03 	&	44.53 	&	0.0382	&	44.54 	&	45.31 	&	45.16 	\\
J0127.9+4857	&	21.9777	&	48.9536	&	LBL	&	0.065	&	7.23 	&	42.92 	&	0.206	&	44.18 	&	43.95 	&	43.65 	\\
J0137.9+5814	&	24.4957	&	58.2494	&	HBL	&	0.275	&	9.55 	&	44.32 	&	0.171	&	44.65 	&	45.18 	&	44.78 	\\
J0139.0+2601	&	24.7581	&	26.0298	&	IBL	&	0.347	&	8.10 	&	44.25 	&	0.129	&	44.70 	&	44.78 	&	44.37 	\\
J0146.9-5202	&	26.729	&	-52.0477	&	IBL	&	0.098	&	9.17 	&	43.41 	&	1.07	&	44.56 	&	43.63 	&	43.28 	\\
J0148.2+5201	&	27.0594	&	52.0243	&	HBL	&	0.437	&	9.75 	&	44.74 	&	0.0444	&	44.63 	&	45.77 	&	45.15 	\\
J0151.3+8601	&	27.8381	&	86.0194	&	IBL	&	0.15	&	9.25 	&	43.54 	&	0.0959	&	44.34 	&	44.13 	&	44.06 	\\
J0152.6+0147	&	28.1614	&	1.7894	&	HBL	&	0.08	&	9.34 	&	42.92 	&	0.0619	&	44.06 	&	44.24 	&	43.54 	\\
J0201.1+0036	&	30.2779	&	0.6029	&	HBL	&	0.298	&	8.19 	&	43.93 	&	0.0132	&	44.29 	&	44.88 	&	44.03 	\\
J0203.7+3042	&	30.9327	&	30.7139	&	LBL	&	0.761	&	8.41 	&	45.01 	&	0.175	&	45.10 	&	46.32 	&	46.69 	\\
J0204.0-3334	&	31.0238	&	-33.5731	&	HBL	&	0.617	&	9.29 	&	44.34 	&	0.0063	&	44.49 	&	45.84 	&	45.06 	\\
J0209.9+7229	&	32.4979	&	72.4877	&	LBL	&	0.895	&	7.76 	&	45.12 	&	0.67	&	45.39 	&	46.54 	&	46.89 	\\
J0219.1-1724	&	34.7821	&	-17.402	&	HBL	&	0.128	&	8.61 	&	43.35 	&	0.0625	&	44.22 	&	44.10 	&	43.36 	\\
J0227.3+0201	&	36.8296	&	2.0203	&	HBL	&	0.457	&	9.47 	&	44.52 	&	0.0368	&	44.62 	&	45.25 	&	45.12 	\\
J0232.8+2018	&	38.2139	&	20.3159	&	HBL	&	0.139	&	10.08 	&	43.18 	&	0.0834	&	44.29 	&	44.60 	&	43.81 	\\
J0237.6-3602	&	39.4244	&	-36.0422	&	HBL	&	0.411	&	8.31 	&	44.66 	&	0.0271	&	44.53 	&	45.43 	&	44.88 	\\
J0238.6+1637	&	39.668	&	16.6179	&	LBL	&	0.94	&	8.58 	&	45.30 	&	1.94	&	45.58 	&	47.45 	&	47.88 	\\
J0250.6+1712	&	42.6563	&	17.2081	&	HBL	&	0.243	&	9.47 	&	44.23 	&	0.07	&	44.47 	&	44.84 	&	44.51 	\\
J0304.5-0054	&	46.1423	&	-0.9148	&	HBL	&	0.511	&	9.16 	&	44.68 	&	0.0233	&	44.60 	&	45.34 	&	44.77 	\\
J0305.1-1608	&	46.2919	&	-16.1466	&	HBL	&	0.311	&	9.24 	&	44.10 	&	2.71	&	45.12 	&	44.68 	&	44.56 	\\
J0325.5-5635	&	51.3794	&	-56.591	&	HBL	&	0.06	&	9.10 	&	42.84 	&	0.0736	&	44.00 	&	43.42 	&	42.90 	\\
J0326.2+0225	&	51.5724	&	2.4228	&	HBL	&	0.147	&	9.21 	&	43.51 	&	0.0682	&	44.28 	&	44.42 	&	44.09 	\\
J0334.2-4008	&	53.5566	&	-40.145	&	LBL	&	1.359	&	8.67 	&	44.66 	&	1.92	&	45.81 	&	47.85 	&	48.10 	\\
J0338.1-2443	&	54.5305	&	-24.7207	&	HBL	&	0.251	&	9.72 	&	43.73 	&	0.0139	&	44.23 	&	44.46 	&	43.67 	\\
J0339.2-1736	&	54.8119	&	-17.6003	&	HBL	&	0.066	&	8.98 	&	43.39 	&	0.171	&	44.15 	&	43.70 	&	43.17 	\\
J0403.5-2437	&	60.8989	&	-24.6168	&	LBL	&	0.357	&	9.75 	&	43.84 	&	0.167	&	44.75 	&	45.07 	&	44.88 	\\
J0407.5+0741	&	61.8921	&	7.6998	&	LBL	&	1.139	&	8.84 	&	45.33 	&	0.296	&	45.41 	&	46.89 	&	47.32 	\\
J0416.9+0105	&	64.2269	&	1.088	&	HBL	&	0.287	&	8.40 	&	44.15 	&	0.121	&	44.61 	&	45.64 	&	44.67 	\\
J0420.2+4012	&	65.0547	&	40.201	&	HBL	&	0.132	&	9.35 	&	43.77 	&	0.116	&	44.32 	&	43.85 	&	43.46 	\\
J0428.6-3756	&	67.173	&	-37.9403	&	LBL	&	1.11	&	8.77 	&	44.90 	&	0.737	&	45.53 	&	47.52 	&	48.32 	\\
J0433.6+2905	&	68.4107	&	29.0975	&	LBL	&	0.91	&	8.25 	&	44.93 	&	0.154	&	45.18 	&	46.02 	&	47.12 	\\
J0505.8-3817	&	76.4749	&	-38.2965	&	HBL	&	0.182	&	9.01 	&	43.91 	&	0.019	&	44.16 	&	44.06 	&	43.63 	\\
J0516.7-6207	&	79.1798	&	-62.1248	&	LBL	&	1.3	&	8.67 	&	45.17 	&	0.418	&	45.55 	&	47.47 	&	47.77 	\\
J0538.8-4405	&	84.7089	&	-44.0862	&	LBL	&	0.896	&	8.45 	&	46.22 	&	2.89	&	45.62 	&	47.81 	&	48.01 	\\
J0558.0-3837	&	89.5233	&	-38.6317	&	HBL	&	0.302	&	9.81 	&	44.52 	&	0.106	&	44.61 	&	45.56 	&	44.80 	\\
J0629.3-1959	&	97.3478	&	-19.9999	&	LBL	&	1.718	&	9.88 	&	45.83 	&	0.677	&	45.82 	&	48.36 	&	48.37 	\\
J0640.0-1253	&	100.0213	&	-12.896	&	HBL	&	0.137	&	9.91 	&	44.02 	&	0.225	&	44.44 	&	44.74 	&	43.96 	\\
J0648.7+1516	&	102.1905	&	15.2808	&	HBL	&	0.179	&	8.84 	&	44.12 	&	0.0647	&	44.34 	&	44.97 	&	44.79 	\\
J0654.7+4246	&	103.6856	&	42.7791	&	LBL	&	0.129	&	7.55 	&	43.02 	&	0.203	&	44.40 	&	43.77 	&	43.75 	\\
			\hline
		\end{tabular}
	\end{minipage}
\end{table*}
\begin{table*}
	\begin{minipage}{150mm}
		\centering
		\contcaption{.}
		\begin{tabular}{@{}crcccccccccccccccrl@{}}
			\hline\hline
			4FGL name & RA & DEC & Type & Redshift & $\rm{\log M}$ & $\log L_{\rm disk}$ & $f_{\nu}$ & $\log P_{\rm kin}$ & $\log L_{\rm sy}$ & $\log L_{\rm ic}$ \\
            {(1)} & {(2)} & {(3)} & {(4)} & {(5)} & {(6)} & {(7)} & {(8)} & {(9)} & {(10)} & {(11)}\\
			\hline
J0656.3+4235	&	104.0918	&	42.5936	&	HBL	&	0.059	&	9.86 	&	42.93 	&	0.934	&	44.38 	&	43.19 	&	42.55 	\\
J0709.1+2241	&	107.2769	&	22.6847	&	HBL	&	0.297	&	9.72 	&	44.39 	&	0.0406	&	44.46 	&	45.07 	&	44.85 	\\
J0710.4+5908	&	107.6234	&	59.1352	&	HBL	&	0.125	&	9.75 	&	42.75 	&	0.159	&	44.35 	&	44.73 	&	44.08 	\\
J0710.9+4733	&	107.7323	&	47.553	&	LBL	&	1.292	&	8.81 	&	45.70 	&	1.02	&	45.68 	&	47.01 	&	47.73 	\\
J0731.9+2805	&	112.981	&	28.0882	&	HBL	&	0.248	&	8.58 	&	43.92 	&	0.0721	&	44.48 	&	44.56 	&	43.94 	\\
J0740.9+3203	&	115.2335	&	32.0581	&	IBL	&	0.179	&	8.46 	&	43.53 	&	0.077	&	44.37 	&	44.04 	&	43.61 	\\
J0744.1+7434	&	116.0345	&	74.5778	&	HBL	&	0.315	&	9.94 	&	44.15 	&	0.0233	&	44.40 	&	45.40 	&	44.78 	\\
J0749.2+2314	&	117.3217	&	23.2337	&	HBL	&	0.174	&	8.71 	&	43.36 	&	0.0547	&	44.31 	&	43.99 	&	43.47 	\\
J0758.9+2703	&	119.7258	&	27.0653	&	IBL	&	0.099	&	8.50 	&	43.08 	&	0.0691	&	44.15 	&	43.50 	&	43.39 	\\
J0803.2-0337	&	120.8247	&	-3.6189	&	HBL	&	0.365	&	6.96 	&	43.95 	&	0.232	&	44.81 	&	44.91 	&	45.24 	\\
J0809.6+3455	&	122.4215	&	34.9252	&	HBL	&	0.082	&	8.80 	&	43.00 	&	0.223	&	44.26 	&	43.83 	&	42.99 	\\
J0809.8+5218	&	122.4617	&	52.3143	&	HBL	&	0.138	&	8.55 	&	43.85 	&	0.183	&	44.41 	&	44.96 	&	44.63 	\\
J0811.4+0146	&	122.861	&	1.7756	&	LBL	&	1.148	&	8.71 	&	44.98 	&	0.599	&	45.52 	&	47.09 	&	47.43 	\\
J0812.0+0237	&	123.0094	&	2.6285	&	HBL	&	0.173	&	9.66 	&	43.57 	&	0.123	&	44.43 	&	44.21 	&	44.04 	\\
J0814.4+2941	&	123.6104	&	29.6857	&	LBL	&	0.374	&	8.58 	&	44.96 	&	0.0047	&	44.22 	&	44.54 	&	44.81 	\\
J0818.4+2816	&	124.6076	&	28.2738	&	IBL	&	0.225	&	8.09 	&	43.97 	&	0.064	&	44.42 	&	44.50 	&	44.02 	\\
J0820.9+2353	&	125.2255	&	23.89	&	HBL	&	0.402	&	8.46 	&	44.26 	&	0.0487	&	44.61 	&	44.99 	&	44.88 	\\
J0823.3+2224	&	125.8443	&	22.4093	&	LBL	&	0.951	&	7.96 	&	44.13 	&	2.27	&	45.61 	&	46.70 	&	46.37 	\\
J0828.3+4152	&	127.0877	&	41.879	&	HBL	&	0.154	&	8.58 	&	43.68 	&	0.091	&	44.34 	&	44.05 	&	43.36 	\\
J0829.0+1755	&	127.2747	&	17.9233	&	HBL	&	0.089	&	8.38 	&	42.92 	&	0.336	&	44.35 	&	43.62 	&	42.91 	\\
J0831.8+0429	&	127.9732	&	4.4941	&	LBL	&	0.174	&	7.00 	&	43.48 	&	1.24	&	44.78 	&	45.10 	&	45.01 	\\
J0832.4+4912	&	128.1078	&	49.2127	&	LBL	&	0.548	&	8.07 	&	44.17 	&	0.344	&	45.04 	&	45.88 	&	45.92 	\\
J0837.3+1458	&	129.3461	&	14.9677	&	HBL	&	0.152	&	7.74 	&	43.43 	&	0.0541	&	44.26 	&	44.65 	&	43.30 	\\
J0842.5+0251	&	130.6331	&	2.8662	&	HBL	&	0.425	&	8.27 	&	44.32 	&	0.0168	&	44.47 	&	44.94 	&	44.56 	\\
J0847.2+1134	&	131.8119	&	11.5692	&	HBL	&	0.198	&	8.04 	&	43.75 	&	0.0332	&	44.28 	&	44.95 	&	44.44 	\\
J0850.5+3455	&	132.6378	&	34.9285	&	HBL	&	0.145	&	8.67 	&	43.82 	&	0.0312	&	44.16 	&	43.89 	&	43.54 	\\
J0854.0+2753	&	133.5155	&	27.884	&	HBL	&	0.494	&	8.52 	&	44.32 	&	0.0144	&	44.51 	&	44.92 	&	44.08 	\\
J0901.4+4542	&	135.3691	&	45.707	&	IBL	&	0.288	&	7.83 	&	43.89 	&	0.021	&	44.35 	&	44.40 	&	44.10 	\\
J0909.7+3104	&	137.4477	&	31.0818	&	HBL	&	0.272	&	8.91 	&	43.96 	&	0.196	&	44.66 	&	44.85 	&	43.98 	\\
J0910.8+3859	&	137.7091	&	38.9999	&	IBL	&	0.199	&	7.73 	&	43.76 	&	0.0103	&	44.10 	&	44.41 	&	43.88 	\\
J0912.9-2102	&	138.2274	&	-21.0446	&	HBL	&	0.198	&	9.53 	&	43.93 	&	0.329	&	44.63 	&	45.16 	&	44.53 	\\
J0916.7+5238	&	139.1906	&	52.6454	&	IBL	&	0.19	&	8.60 	&	43.66 	&	0.139	&	44.48 	&	44.44 	&	43.76 	\\
J0917.3-0342	&	139.3339	&	-3.7035	&	HBL	&	0.308	&	9.34 	&	44.09 	&	0.0321	&	44.44 	&	44.81 	&	44.25 	\\
J0930.5+4951	&	142.6254	&	49.8577	&	HBL	&	0.187	&	8.87 	&	43.64 	&	0.0214	&	44.19 	&	44.73 	&	43.90 	\\
J0932.7+1041	&	143.1802	&	10.6903	&	HBL	&	0.361	&	9.16 	&	44.12 	&	0.0336	&	44.51 	&	44.90 	&	44.53 	\\
J0940.4+6148	&	145.1207	&	61.8156	&	HBL	&	0.211	&	8.63 	&	43.66 	&	0.0128	&	44.15 	&	44.43 	&	43.96 	\\
J0942.3+2842	&	145.5806	&	28.7091	&	HBL	&	0.366	&	8.60 	&	43.92 	&	0.022	&	44.45 	&	44.25 	&	44.18 	\\
J0945.7+5759	&	146.432	&	57.9871	&	IBL	&	0.229	&	8.62 	&	43.90 	&	0.112	&	44.51 	&	44.64 	&	44.15 	\\
J0946.2+0104	&	146.5672	&	1.0701	&	HBL	&	0.128	&	7.70 	&	43.43 	&	0.154	&	44.36 	&	43.88 	&	43.45 	\\
J0955.1+3551	&	148.7816	&	35.8584	&	HBL	&	0.557	&	8.92 	&	44.44 	&	0.0078	&	44.47 	&	45.53 	&	44.67 	\\
J0959.4+2120	&	149.8712	&	21.3459	&	HBL	&	0.365	&	8.28 	&	44.16 	&	0.0408	&	44.54 	&	45.27 	&	44.48 	\\
J1001.1+2911	&	150.2938	&	29.188	&	LBL	&	0.556	&	7.78 	&	44.68 	&	0.146	&	44.92 	&	46.01 	&	45.95 	\\
J1010.2-3119	&	152.5716	&	-31.3207	&	HBL	&	0.143	&	9.97 	&	43.57 	&	0.0743	&	44.28 	&	44.66 	&	44.02 	\\
J1023.8+3002	&	155.9608	&	30.0458	&	HBL	&	0.433	&	9.06 	&	44.43 	&	0.009	&	44.38 	&	45.01 	&	44.75 	\\
J1026.9+0608	&	156.7322	&	6.1431	&	HBL	&	0.449	&	9.20 	&	44.40 	&	0.0105	&	44.42 	&	45.28 	&	44.75 	\\
J1031.1+7442	&	157.7925	&	74.7019	&	LBL	&	0.123	&	7.87 	&	43.88 	&	0.209	&	44.39 	&	44.18 	&	44.22 	\\
J1031.3+5053	&	157.8454	&	50.8839	&	HBL	&	0.36	&	8.07 	&	44.14 	&	0.0379	&	44.52 	&	45.83 	&	45.41 	\\
J1033.5+4221	&	158.3827	&	42.3507	&	HBL	&	0.211	&	8.65 	&	43.52 	&	0.0448	&	44.35 	&	43.90 	&	43.41 	\\
J1035.6+4409	&	158.9225	&	44.1609	&	IBL	&	0.444	&	7.76 	&	44.06 	&	0.0429	&	44.63 	&	44.47 	&	45.02 	\\
J1041.7+3902	&	160.4411	&	39.0426	&	HBL	&	0.208	&	8.73 	&	43.74 	&	0.0223	&	44.23 	&	44.05 	&	43.62 	\\
J1041.9-0557	&	160.4774	&	-5.9528	&	HBL	&	0.39	&	8.54 	&	44.18 	&	0.0836	&	44.68 	&	45.24 	&	44.70 	\\
J1043.2+2408	&	160.8053	&	24.146	&	LBL	&	0.559	&	8.63 	&	44.89 	&	0.325	&	45.04 	&	46.02 	&	45.94 	\\
J1046.8-2534	&	161.7027	&	-25.5749	&	HBL	&	0.254	&	9.94 	&	43.77 	&	0.0141	&	44.24 	&	44.97 	&	44.25 	\\
J1049.5+1548	&	162.3892	&	15.8086	&	IBL	&	0.326	&	8.25 	&	44.41 	&	0.0506	&	44.53 	&	45.09 	&	44.86 	\\
J1049.7+5011	&	162.4334	&	50.1836	&	HBL	&	0.402	&	8.49 	&	44.01 	&	0.0078	&	44.33 	&	44.80 	&	44.45 	\\
J1051.4+3942	&	162.8702	&	39.7159	&	HBL	&	0.497	&	9.10 	&	44.48 	&	0.0108	&	44.47 	&	45.56 	&	45.03 	\\
J1051.9+0103	&	162.9886	&	1.0647	&	IBL	&	0.265	&	7.61 	&	43.69 	&	0.0163	&	44.28 	&	44.14 	&	43.80 	\\
J1053.7+4930	&	163.4253	&	49.5081	&	HBL	&	0.14	&	8.74 	&	43.46 	&	0.0646	&	44.26 	&	44.16 	&	43.61 	\\
J1057.8-2754	&	164.454	&	-27.9016	&	HBL	&	0.091	&	9.36 	&	43.69 	&	0.0638	&	44.11 	&	43.74 	&	43.06 	\\
J1058.4+0133	&	164.624	&	1.5641	&	LBL	&	0.892	&	9.50 	&	45.74 	&	3.22	&	45.63 	&	47.26 	&	47.48 	\\ 
 			\hline
		\end{tabular}
	\end{minipage}
\end{table*}
\begin{table*}
	\begin{minipage}{150mm}
		\centering
		\contcaption{.}
		\begin{tabular}{@{}crcccccccccccccccrl@{}}
			\hline\hline
			4FGL name & RA & DEC & Type & Redshift & $\rm{\log M}$ & $\log L_{\rm disk}$ & $f_{\nu}$ & $\log P_{\rm kin}$ & $\log L_{\rm sy}$ & $\log L_{\rm ic}$ \\
           {(1)} & {(2)} & {(3)} & {(4)} & {(5)} & {(6)} & {(7)} & {(8)} & {(9)} & {(10)} & {(11)}\\
			\hline
J1058.6+5627	&	164.6652	&	56.4634	&	HBL	&	0.143	&	8.33 	&	43.90 	&	0.229	&	44.46 	&	44.81 	&	44.55 	\\
J1058.6-8003	&	164.66	&	-80.064	&	LBL	&	0.581	&	8.64 	&	45.49 	&	0.77	&	45.19 	&	46.37 	&	46.35 	\\
J1104.4+3812	&	166.1187	&	38.207	&	HBL	&	0.033	&	9.05 	&	42.18 	&	0.769	&	44.16 	&	44.62 	&	44.51 	\\
J1105.8+3944	&	166.4589	&	39.7426	&	LBL	&	0.099	&	8.29 	&	42.67 	&	0.045	&	44.08 	&	43.71 	&	42.86 	\\
J1109.6+3735	&	167.4092	&	37.5868	&	HBL	&	0.398	&	8.96 	&	44.16 	&	0.0044	&	44.24 	&	44.79 	&	44.59 	\\
J1112.4+1751	&	168.1132	&	17.8509	&	HBL	&	0.421	&	8.46 	&	44.45 	&	0.0143	&	44.44 	&	45.24 	&	44.73 	\\
J1117.0+2013	&	169.2708	&	20.2294	&	HBL	&	0.138	&	8.51 	&	43.71 	&	0.103	&	44.32 	&	44.52 	&	44.43 	\\
J1119.6-3047	&	169.9245	&	-30.7962	&	HBL	&	0.412	&	9.75 	&	44.03 	&	0.0094	&	44.37 	&	44.84 	&	44.48 	\\
J1124.4+2308	&	171.102	&	23.137	&	LBL	&	0.795	&	7.86 	&	44.87 	&	0.154	&	45.10 	&	45.55 	&	45.88 	\\
J1125.5-3557	&	171.3929	&	-35.9581	&	LBL	&	0.284	&	7.28 	&	43.55 	&	0.208	&	44.69 	&	44.87 	&	44.82 	\\
J1130.5-3137	&	172.6499	&	-31.6219	&	HBL	&	0.151	&	9.24 	&	43.43 	&	0.0265	&	44.15 	&	43.92 	&	43.64 	\\
J1130.8+1016	&	172.7192	&	10.2728	&	IBL	&	0.172	&	8.41 	&	44.01 	&	0.0136	&	44.09 	&	44.23 	&	44.11 	\\
J1131.4+5809	&	172.8555	&	58.151	&	IBL	&	0.36	&	8.42 	&	44.21 	&	0.0444	&	44.55 	&	45.19 	&	44.94 	\\
J1136.4+6736	&	174.1179	&	67.6127	&	HBL	&	0.134	&	8.67 	&	43.49 	&	0.0458	&	44.19 	&	44.56 	&	44.10 	\\
J1136.8+2550	&	174.2171	&	25.8463	&	HBL	&	0.154	&	9.46 	&	43.53 	&	0.0163	&	44.08 	&	44.31 	&	43.20 	\\
J1140.5+1528	&	175.129	&	15.4824	&	HBL	&	0.244	&	8.70 	&	43.97 	&	0.07	&	44.47 	&	44.98 	&	44.02 	\\
J1147.0-3812	&	176.76	&	-38.2006	&	LBL	&	1.053	&	8.69 	&	45.38 	&	1.8	&	45.64 	&	47.27 	&	47.31 	\\
J1149.4+2441	&	177.3713	&	24.6873	&	HBL	&	0.402	&	8.21 	&	44.32 	&	0.00758	&	44.33 	&	45.70 	&	44.62 	\\
J1152.1+2837	&	178.0309	&	28.6293	&	HBL	&	0.441	&	9.06 	&	44.37 	&	0.0246	&	44.54 	&	45.20 	&	44.67 	\\
J1153.7+3822	&	178.4464	&	38.3684	&	LBL	&	0.41	&	8.11 	&	44.26 	&	0.15	&	44.79 	&	45.08 	&	44.31 	\\
J1154.0-0010	&	178.5103	&	-0.1787	&	HBL	&	0.254	&	8.13 	&	43.75 	&	0.0106	&	44.19 	&	44.65 	&	44.19 	\\
J1202.4+4442	&	180.61	&	44.7147	&	IBL	&	0.297	&	8.80 	&	44.11 	&	0.106	&	44.60 	&	44.40 	&	44.24 	\\
J1203.1+6031	&	180.7881	&	60.518	&	IBL	&	0.065	&	8.40 	&	43.20 	&	0.191	&	44.17 	&	43.84 	&	43.18 	\\
J1203.4-3925	&	180.852	&	-39.4257	&	HBL	&	0.227	&	8.88 	&	43.64 	&	0.0646	&	44.43 	&	44.24 	&	44.22 	\\
J1204.0+1146	&	181.0204	&	11.7761	&	HBL	&	0.296	&	8.37 	&	43.89 	&	0.0151	&	44.31 	&	45.01 	&	44.46 	\\
J1212.0+2242	&	183.0146	&	22.709	&	HBL	&	0.453	&	8.16 	&	44.50 	&	0.0202	&	44.53 	&	45.48 	&	44.52 	\\
J1215.1+0731	&	183.7895	&	7.5222	&	HBL	&	0.136	&	8.89 	&	43.49 	&	0.138	&	44.36 	&	44.04 	&	43.48 	\\
J1216.1+0930	&	184.0415	&	9.5096	&	HBL	&	0.094	&	8.95 	&	42.97 	&	0.209	&	44.30 	&	43.71 	&	43.18 	\\
J1219.7-0313	&	184.9334	&	-3.2217	&	HBL	&	0.299	&	8.45 	&	44.18 	&	0.0289	&	44.41 	&	44.77 	&	44.63 	\\
J1221.3+3010	&	185.3449	&	30.1677	&	HBL	&	0.184	&	9.00 	&	44.17 	&	0.0715	&	44.37 	&	45.19 	&	45.10 	\\
J1221.5+2814	&	185.3784	&	28.2382	&	IBL	&	0.102	&	8.55 	&	43.81 	&	0.732	&	44.52 	&	44.71 	&	44.58 	\\
J1223.9+7954	&	185.9854	&	79.9025	&	HBL	&	0.375	&	9.53 	&	44.32 	&	0.0315	&	44.51 	&	44.44 	&	44.15 	\\
J1224.4+2436	&	186.1161	&	24.6142	&	HBL	&	0.219	&	8.29 	&	44.11 	&	0.0259	&	44.28 	&	45.10 	&	44.62 	\\
J1231.5+1421	&	187.8771	&	14.3532	&	IBL	&	0.256	&	8.73 	&	44.07 	&	0.05	&	44.43 	&	44.59 	&	44.38 	\\
J1231.6+6415	&	187.9012	&	64.2535	&	HBL	&	0.163	&	8.96 	&	43.49 	&	0.0588	&	44.29 	&	44.26 	&	43.80 	\\
J1233.6+5027	&	188.4105	&	50.4606	&	IBL	&	0.207	&	8.32 	&	43.59 	&	0.283	&	44.62 	&	44.27 	&	44.04 	\\
J1236.3+3858	&	189.0894	&	38.9814	&	IBL	&	0.389	&	8.54 	&	44.15 	&	0.037	&	44.55 	&	44.79 	&	44.49 	\\
J1237.8+6256	&	189.457	&	62.9342	&	HBL	&	0.297	&	8.64 	&	43.86 	&	0.0125	&	44.28 	&	44.60 	&	43.84 	\\
J1244.5+1616	&	191.1361	&	16.2803	&	HBL	&	0.456	&	8.76 	&	44.23 	&	0.154	&	44.84 	&	44.78 	&	44.77 	\\
J1246.3+0112	&	191.5786	&	1.2166	&	IBL	&	0.386	&	7.57 	&	44.14 	&	0.045	&	44.58 	&	44.63 	&	44.39 	\\
J1248.7+5127	&	192.1805	&	51.463	&	IBL	&	0.351	&	8.24 	&	44.21 	&	0.115	&	44.68 	&	44.98 	&	44.46 	\\
J1250.6+0217	&	192.6513	&	2.2876	&	LBL	&	0.954	&	8.69 	&	45.08 	&	0.341	&	45.33 	&	46.22 	&	46.43 	\\
J1251.2+1039	&	192.821	&	10.6536	&	IBL	&	0.245	&	7.62 	&	43.89 	&	0.154	&	44.59 	&	44.44 	&	44.10 	\\
J1254.9-4426	&	193.728	&	-44.4441	&	LBL	&	0.041	&	8.95 	&	42.91 	&	0.368	&	44.12 	&	42.78 	&	42.70 	\\
J1256.2-1146	&	194.0632	&	-11.7755	&	HBL	&	0.058	&	8.94 	&	43.06 	&	0.0544	&	43.94 	&	43.53 	&	42.99 	\\
J1257.6+2413	&	194.4191	&	24.2199	&	HBL	&	0.141	&	8.56 	&	43.33 	&	0.0147	&	44.03 	&	44.19 	&	43.15 	\\
J1258.3+6121	&	194.5879	&	61.3622	&	HBL	&	0.224	&	7.90 	&	43.91 	&	0.0102	&	44.14 	&	43.95 	&	43.77 	\\
J1319.5+1404	&	199.8979	&	14.0708	&	IBL	&	0.573	&	8.31 	&	44.94 	&	0.077	&	44.84 	&	45.64 	&	45.26 	\\
J1321.9+3219	&	200.4801	&	32.3313	&	IBL	&	0.396	&	8.73 	&	43.90 	&	0.0164	&	44.44 	&	45.00 	&	44.53 	\\
J1322.9+0437	&	200.7372	&	4.6308	&	HBL	&	0.224	&	8.94 	&	43.70 	&	0.037	&	44.34 	&	44.36 	&	43.93 	\\
J1326.1+1232	&	201.5493	&	12.5348	&	HBL	&	0.204	&	8.75 	&	43.70 	&	0.0574	&	44.37 	&	44.50 	&	43.72 	\\
J1331.0+5653	&	202.7577	&	56.894	&	HBL	&	0.27	&	8.63 	&	43.75 	&	0.00516	&	44.11 	&	44.50 	&	43.91 	\\
J1331.2-1325	&	202.8192	&	-13.4282	&	LBL	&	0.251	&	7.80 	&	43.52 	&	0.0512	&	44.43 	&	44.26 	&	45.01 	\\
J1335.3-2949	&	203.8475	&	-29.8295	&	HBL	&	0.513	&	9.78 	&	44.32 	&	0.0106	&	44.48 	&	45.76 	&	45.05 	\\
J1336.2+2320	&	204.051	&	23.3349	&	HBL	&	0.267	&	7.64 	&	43.76 	&	0.0134	&	44.25 	&	44.46 	&	44.02 	\\
J1340.8-0409	&	205.2171	&	-4.1612	&	HBL	&	0.223	&	9.60 	&	43.96 	&	0.027	&	44.29 	&	44.46 	&	44.13 	\\
J1341.2+3958	&	205.3209	&	39.9738	&	HBL	&	0.171	&	8.91 	&	43.45 	&	0.0878	&	44.37 	&	44.52 	&	43.71 	\\
J1341.6+5515	&	205.4087	&	55.254	&	HBL	&	0.207	&	8.17 	&	43.56 	&	0.0379	&	44.31 	&	44.06 	&	43.09 	\\
J1342.7+0505	&	205.6851	&	5.0904	&	IBL	&	0.136	&	8.56 	&	43.37 	&	1.6	&	44.73 	&	44.57 	&	43.90 	\\
J1351.9+2847	&	207.9959	&	28.7964	&	HBL	&	0.268	&	8.55 	&	43.76 	&	0.011	&	44.22 	&	44.05 	&	43.62 	\\
			\hline
		\end{tabular}
	\end{minipage}
\end{table*}
\begin{table*}
	\begin{minipage}{150mm}
		\centering
		\contcaption{.}
		\begin{tabular}{@{}crcccccccccccccccrl@{}}
			\hline\hline
			4FGL name & RA & DEC & Type & Redshift & $\rm{\log M}$ & $\log L_{\rm disk}$ & $f_{\nu}$ & $\log P_{\rm kin}$ & $\log L_{\rm sy}$ & $\log L_{\rm ic}$ \\
			{(1)} & {(2)} & {(3)} & {(4)} & {(5)} & {(6)} & {(7)} & {(8)} & {(9)} & {(10)} & {(11)}\\
			\hline
J1353.2+3740	&	208.3049	&	37.682	&	HBL	&	0.216	&	8.99 	&	43.61 	&	0.0343	&	44.31 	&	44.35 	&	43.81 	\\
J1353.4+5600	&	208.3602	&	56.0024	&	HBL	&	0.404	&	8.29 	&	44.26 	&	0.0149	&	44.43 	&	44.89 	&	44.30 	\\
J1402.6+1600	&	210.6584	&	16.0016	&	IBL	&	0.245	&	7.30 	&	43.66 	&	0.85	&	44.85 	&	44.64 	&	43.92 	\\
J1403.4+4319	&	210.8684	&	43.3225	&	HBL	&	0.493	&	8.16 	&	44.27 	&	0.0205	&	44.57 	&	44.74 	&	44.65 	\\
J1404.8+6554	&	211.2158	&	65.9048	&	HBL	&	0.363	&	8.41 	&	44.26 	&	0.0154	&	44.39 	&	44.92 	&	44.66 	\\
J1406.9+1643	&	211.742	&	16.7206	&	HBL	&	0.603	&	8.90 	&	44.77 	&	0.0083	&	44.52 	&	45.89 	&	45.28 	\\
J1410.3+1438	&	212.5908	&	14.6434	&	IBL	&	0.144	&	8.37 	&	43.33 	&	0.494	&	44.57 	&	43.98 	&	43.65 	\\
J1411.8+5249	&	212.9692	&	52.8278	&	HBL	&	0.076	&	8.82 	&	42.78 	&	0.848	&	44.44 	&	43.43 	&	42.63 	\\
J1412.1+7427	&	213.0383	&	74.45	&	IBL	&	0.436	&	6.35 	&	43.45 	&	0.107	&	44.76 	&	44.93 	&	45.28 	\\
J1415.5+4830	&	213.8992	&	48.5142	&	HBL	&	0.496	&	7.73 	&	44.65 	&	0.0287	&	44.62 	&	45.27 	&	45.03 	\\
J1416.1-2417	&	214.0334	&	-24.2982	&	HBL	&	0.136	&	9.21 	&	43.48 	&	0.0729	&	44.26 	&	44.43 	&	43.45 	\\
J1417.9+2543	&	214.494	&	25.7238	&	HBL	&	0.236	&	8.17 	&	43.59 	&	0.0887	&	44.49 	&	45.30 	&	44.42 	\\
J1419.8+5423	&	214.955	&	54.3937	&	LBL	&	0.152	&	9.12 	&	44.17 	&	0.789	&	44.66 	&	45.15 	&	44.42 	\\
J1424.1-1750	&	216.0294	&	-17.8447	&	HBL	&	0.082	&	8.59 	&	43.27 	&	0.013	&	43.83 	&	43.43 	&	42.90 	\\
J1428.5+4240	&	217.1286	&	42.6776	&	HBL	&	0.129	&	8.59 	&	43.59 	&	0.032	&	44.12 	&	44.88 	&	44.39 	\\
J1439.3+3932	&	219.8299	&	39.538	&	HBL	&	0.344	&	8.79 	&	44.65 	&	0.0442	&	44.53 	&	45.50 	&	44.86 	\\
J1440.9+0609	&	220.242	&	6.1631	&	HBL	&	0.396	&	9.49 	&	43.35 	&	0.0902	&	44.70 	&	45.42 	&	45.16 	\\
J1442.6-4623	&	220.6605	&	-46.3869	&	HBL	&	0.103	&	9.25 	&	43.25 	&	0.0964	&	44.21 	&	44.13 	&	43.38 	\\
J1442.7+1200	&	220.698	&	12.0126	&	HBL	&	0.163	&	8.74 	&	43.64 	&	0.06	&	44.30 	&	44.77 	&	44.06 	\\
J1443.6+2515	&	220.9028	&	25.2631	&	HBL	&	0.529	&	8.13 	&	44.49 	&	0.0067	&	44.43 	&	45.21 	&	44.94 	\\
J1503.5+4759	&	225.8955	&	47.9959	&	IBL	&	0.345	&	7.97 	&	44.51 	&	0.106	&	44.66 	&	45.08 	&	45.51 	\\
J1506.4+4331	&	226.6221	&	43.5257	&	IBL	&	0.47	&	9.03 	&	44.37 	&	0.0291	&	44.60 	&	44.85 	&	44.67 	\\
J1507.2+1721	&	226.8207	&	17.3519	&	HBL	&	0.565	&	8.19 	&	44.48 	&	0.0234	&	44.65 	&	45.36 	&	45.34 	\\
J1508.8+2708	&	227.2045	&	27.1407	&	HBL	&	0.27	&	8.30 	&	43.95 	&	0.0374	&	44.41 	&	45.26 	&	44.32 	\\
J1516.8+2918	&	229.2126	&	29.3123	&	IBL	&	0.13	&	8.79 	&	43.24 	&	0.0746	&	44.25 	&	44.18 	&	44.01 	\\
J1517.7-2422	&	229.4254	&	-24.373	&	LBL	&	0.048	&	9.09 	&	43.15 	&	2.04	&	44.43 	&	44.19 	&	43.93 	\\
J1518.6+4044	&	229.6606	&	40.7449	&	HBL	&	0.065	&	8.25 	&	42.65 	&	0.0409	&	43.93 	&	43.25 	&	42.72 	\\
J1523.2+0533	&	230.8241	&	5.5569	&	HBL	&	0.176	&	8.57 	&	43.33 	&	0.0362	&	44.25 	&	44.08 	&	43.42 	\\
J1532.0+3016	&	233.0159	&	30.2685	&	HBL	&	0.065	&	8.27 	&	42.69 	&	0.0544	&	43.97 	&	43.48 	&	42.99 	\\
J1533.2+1855	&	233.3103	&	18.9201	&	HBL	&	0.307	&	8.27 	&	44.07 	&	0.0229	&	44.38 	&	45.12 	&	44.57 	\\
J1534.8+3716	&	233.7218	&	37.2723	&	HBL	&	0.143	&	7.83 	&	43.56 	&	0.0224	&	44.10 	&	44.01 	&	43.60 	\\
J1535.4+3919	&	233.8741	&	39.3194	&	HBL	&	0.257	&	8.31 	&	44.13 	&	0.0197	&	44.29 	&	44.92 	&	44.05 	\\
J1540.7+1449	&	235.1903	&	14.822	&	LBL	&	0.606	&	8.23 	&	44.18 	&	1.39	&	45.30 	&	46.09 	&	45.96 	\\
J1541.7+1413	&	235.4469	&	14.2306	&	HBL	&	0.223	&	7.94 	&	43.80 	&	0.0354	&	44.33 	&	44.75 	&	43.96 	\\
J1548.3+1456	&	237.0999	&	14.9461	&	IBL	&	0.23	&	9.32 	&	43.40 	&	0.0244	&	44.28 	&	44.40 	&	44.81 	\\
J1558.9-6432	&	239.7377	&	-64.5404	&	HBL	&	0.08	&	8.94 	&	43.39 	&	0.94	&	44.48 	&	44.08 	&	43.59 	\\
J1605.5+5423	&	241.3848	&	54.3982	&	HBL	&	0.212	&	7.92 	&	43.43 	&	0.0076	&	44.08 	&	44.31 	&	43.90 	\\
J1606.3+5629	&	241.5917	&	56.4975	&	HBL	&	0.437	&	8.48 	&	44.30 	&	0.0157	&	44.47 	&	45.41 	&	44.80 	\\
J1616.7+4107	&	244.1821	&	41.1234	&	LBL	&	0.267	&	8.67 	&	43.88 	&	0.0958	&	44.55 	&	44.69 	&	44.61 	\\
J1626.3+3514	&	246.582	&	35.2382	&	HBL	&	0.498	&	8.82 	&	44.47 	&	0.0201	&	44.57 	&	45.28 	&	45.05 	\\
J1626.6-7639	&	246.6553	&	-76.6502	&	IBL	&	0.105	&	8.92 	&	43.56 	&	0.0065	&	43.81 	&	43.68 	&	43.43 	\\
J1627.3+3148	&	246.8347	&	31.814	&	HBL	&	0.58	&	8.41 	&	44.47 	&	0.00369	&	44.38 	&	45.12 	&	44.99 	\\
J1637.6+4548	&	249.4137	&	45.8108	&	HBL	&	0.192	&	8.52 	&	43.53 	&	0.019	&	44.18 	&	44.30 	&	43.47 	\\
J1644.2+4546	&	251.0556	&	45.776	&	HBL	&	0.225	&	8.97 	&	43.70 	&	0.184	&	44.58 	&	44.55 	&	43.46 	\\
J1647.5+2911	&	251.8835	&	29.1837	&	HBL	&	0.133	&	8.77 	&	43.18 	&	0.388	&	44.51 	&	43.92 	&	43.27 	\\
J1647.5+4950	&	251.8923	&	49.8336	&	LBL	&	0.047	&	7.98 	&	43.27 	&	0.178	&	44.05 	&	43.65 	&	43.70 	\\
J1653.8+3945	&	253.4738	&	39.7595	&	HBL	&	0.033	&	9.91 	&	43.30 	&	1.56	&	44.27 	&	44.01 	&	43.90 	\\
J1658.4+6150	&	254.62	&	61.8483	&	HBL	&	0.374	&	7.66 	&	44.15 	&	0.0384	&	44.54 	&	44.93 	&	44.60 	\\
J1704.2+1234	&	256.0599	&	12.5752	&	LBL	&	0.452	&	9.45 	&	44.60 	&	0.03	&	44.59 	&	45.20 	&	45.74 	\\
J1730.8+3715	&	262.7026	&	37.2641	&	HBL	&	0.204	&	9.71 	&	43.62 	&	0.0631	&	44.39 	&	44.32 	&	44.06 	\\
J1744.0+1935	&	266.008	&	19.5956	&	HBL	&	0.084	&	9.69 	&	43.28 	&	0.551	&	44.41 	&	44.17 	&	43.53 	\\
J1745.6+3950	&	266.4158	&	39.8412	&	HBL	&	0.267	&	9.52 	&	43.90 	&	0.636	&	44.84 	&	44.51 	&	43.91 	\\
J1751.5+0938	&	267.8776	&	9.6456	&	LBL	&	0.322	&	7.98 	&	44.08 	&	0.623	&	44.91 	&	45.91 	&	45.87 	\\
J1754.5-6425	&	268.639	&	-64.418	&	LBL	&	1.255	&	8.13 	&	44.91 	&	0.177	&	45.40 	&	47.17 	&	47.48 	\\
J1800.6+7828	&	270.173	&	78.4674	&	LBL	&	0.691	&	8.94 	&	45.68 	&	2.22	&	45.44 	&	46.96 	&	46.93 	\\
J1806.8+6949	&	271.7108	&	69.827	&	LBL	&	0.05	&	7.10 	&	43.41 	&	1.89	&	44.43 	&	44.23 	&	43.73 	\\
J1853.8+6714	&	283.4625	&	67.2487	&	HBL	&	0.212	&	8.96 	&	43.56 	&	0.011	&	44.13 	&	44.46 	&	43.58 	\\
J1911.4-1908	&	287.8681	&	-19.1494	&	HBL	&	0.138	&	9.27 	&	43.46 	&	0.44	&	44.54 	&	44.23 	&	43.94 	\\
J1917.7-1921	&	289.4384	&	-19.3628	&	HBL	&	0.137	&	9.08 	&	43.82 	&	0.482	&	44.55 	&	44.69 	&	44.46 	\\
J1942.8-3512	&	295.7173	&	-35.2012	&	IBL	&	0.05	&	9.01 	&	42.52 	&	0.154	&	44.05 	&	42.85 	&	42.96 	\\			
			\hline
		\end{tabular}
	\end{minipage}
\end{table*}
\begin{table*}
	\begin{minipage}{150mm}
	   \centering
		\contcaption{.}
		\begin{tabular}{@{}crcccccccccccccccrl@{}}
			\hline\hline
			4FGL name & RA & DEC & Type & Redshift & $\rm{\log M}$ & $\log L_{\rm disk}$ & $f_{\nu}$ & $\log P_{\rm kin}$ & $\log L_{\rm sy}$ & $\log L_{\rm ic}$ \\
			{(1)} & {(2)} & {(3)} & {(4)} & {(5)} & {(6)} & {(7)} & {(8)} & {(9)} & {(10)} & {(11)}\\
			\hline
J1954.6-1122	&	298.6693	&	-11.3815	&	LBL	&	0.683	&	8.05 	&	44.46 	&	0.367	&	45.16 	&	46.16 	&	46.74 	\\
J2000.0+6508	&	300.011	&	65.1479	&	HBL	&	0.047	&	9.07 	&	42.89 	&	0.25	&	44.10 	&	44.50 	&	43.89 	\\
J2014.3-0047	&	303.599	&	-0.7922	&	HBL	&	0.23	&	9.18 	&	44.31 	&	0.125	&	44.53 	&	44.73 	&	44.34 	\\
J2032.0+1219	&	308.004	&	12.3279	&	LBL	&	1.211	&	8.31 	&	44.94 	&	0.999	&	45.64 	&	47.29 	&	47.51 	\\
J2049.7-0036	&	312.4456	&	-0.616	&	HBL	&	0.257	&	8.00 	&	43.66 	&	0.0058	&	44.11 	&	44.71 	&	43.78 	\\
J2054.8+0015	&	313.7246	&	0.257	&	HBL	&	0.151	&	8.75 	&	43.40 	&	0.057	&	44.26 	&	44.19 	&	43.77 	\\
J2108.7-0250	&	317.1788	&	-2.8449	&	HBL	&	0.149	&	10.10 	&	43.67 	&	0.127	&	44.38 	&	44.29 	&	43.62 	\\
J2115.9-0113	&	318.9959	&	-1.2306	&	HBL	&	0.305	&	8.22 	&	44.21 	&	0.069	&	44.55 	&	44.80 	&	44.53 	\\
J2130.2-7320	&	322.5524	&	-73.3348	&	HBL	&	0.057	&	9.49 	&	43.00 	&	0.182	&	44.12 	&	43.26 	&	42.05 	\\
J2134.2-0154	&	323.5699	&	-1.9042	&	LBL	&	1.283	&	8.75 	&	44.83 	&	1.69	&	45.75 	&	47.54 	&	47.56 	\\
J2150.8+1118	&	327.7033	&	11.3149	&	HBL	&	0.495	&	7.24 	&	44.55 	&	0.0099	&	44.46 	&	45.40 	&	44.94 	\\
J2152.5+1737	&	328.137	&	17.6173	&	LBL	&	0.872	&	8.61 	&	44.41 	&	0.681	&	45.38 	&	46.84 	&	46.66 	\\
J2153.1-0041	&	328.2823	&	-0.6927	&	HBL	&	0.342	&	9.75 	&	44.16 	&	0.0247	&	44.44 	&	45.31 	&	44.64 	\\
J2158.8-3013	&	329.7141	&	-30.2251	&	HBL	&	0.116	&	8.91 	&	43.51 	&	0.49	&	44.50 	&	45.86 	&	45.35 	\\
J2202.7+4216	&	330.6946	&	42.2821	&	LBL	&	0.069	&	7.81 	&	43.34 	&	6.05	&	44.71 	&	45.01 	&	44.69 	\\
J2202.7-5637	&	330.6995	&	-56.6318	&	HBL	&	0.049	&	9.44 	&	42.66 	&	0.0584	&	43.90 	&	42.91 	&	41.94 	\\
J2204.3+0438	&	331.0832	&	4.6401	&	IBL	&	0.027	&	7.32 	&	42.75 	&	0.467	&	44.02 	&	42.87 	&	42.78 	\\
J2206.8-0032	&	331.7087	&	-0.5461	&	LBL	&	1.053	&	8.58 	&	44.64 	&	0.152	&	45.26 	&	46.86 	&	46.93 	\\
J2209.7-0451	&	332.4382	&	-4.8597	&	HBL	&	0.397	&	8.59 	&	44.45 	&	0.0248	&	44.50 	&	45.08 	&	44.57 	\\
J2211.0-0003	&	332.7625	&	-0.0635	&	HBL	&	0.362	&	8.55 	&	44.26 	&	0.0255	&	44.47 	&	45.14 	&	44.37 	\\
J2216.9+2421	&	334.238	&	24.3575	&	LBL	&	1.033	&	8.44 	&	45.04 	&	0.528	&	45.44 	&	47.18 	&	47.09 	\\
J2220.5+2813	&	335.1419	&	28.2322	&	HBL	&	0.149	&	8.62 	&	43.31 	&	0.0488	&	44.23 	&	44.12 	&	43.51 	\\
J2232.8+1334	&	338.2245	&	13.5764	&	HBL	&	0.214	&	8.32 	&	43.55 	&	0.0251	&	44.26 	&	44.98 	&	43.75 	\\
J2243.4-2544	&	340.8654	&	-25.7363	&	LBL	&	0.774	&	8.47 	&	44.51 	&	1.1	&	45.39 	&	46.63 	&	46.57 	\\
J2250.0+3825	&	342.5142	&	38.4247	&	HBL	&	0.119	&	9.44 	&	43.31 	&	0.104	&	44.27 	&	44.56 	&	44.13 	\\
J2252.6+1245	&	343.1676	&	12.7543	&	HBL	&	0.497	&	8.08 	&	44.65 	&	0.0463	&	44.69 	&	45.25 	&	44.63 	\\
J2253.7+1405	&	343.4483	&	14.091	&	HBL	&	0.327	&	8.49 	&	44.08 	&	0.0116	&	44.31 	&	44.98 	&	44.30 	\\
J2257.5+0748	&	344.3874	&	7.8014	&	LBL	&	0.19	&	9.66 	&	43.75 	&	0.394	&	44.64 	&	44.63 	&	43.99 	\\
J2314.0+1445	&	348.5081	&	14.7532	&	HBL	&	0.164	&	9.11 	&	43.55 	&	0.0406	&	44.24 	&	44.57 	&	44.04 	\\
J2315.6-5018	&	348.914	&	-50.3127	&	LBL	&	0.811	&	8.41 	&	44.71 	&	0.233	&	45.18 	&	46.66 	&	46.34 	\\
J2319.1-4207	&	349.7763	&	-42.1173	&	HBL	&	0.055	&	9.41 	&	43.30 	&	1.67	&	44.44 	&	43.55 	&	42.57 	\\
J2322.7+3436	&	350.6849	&	34.6125	&	HBL	&	0.098	&	9.55 	&	42.80 	&	0.0957	&	44.20 	&	44.09 	&	43.28 	\\
J2330.3-2948	&	352.5806	&	-29.8072	&	HBL	&	0.297	&	8.99 	&	44.20 	&	0.039	&	44.45 	&	44.54 	&	44.05 	\\
J2343.6+3438	&	355.9063	&	34.6403	&	HBL	&	0.365	&	8.47 	&	44.19 	&	0.0349	&	44.52 	&	45.72 	&	45.16 	\\
J2346.7+0705	&	356.6786	&	7.0931	&	HBL	&	0.172	&	8.79 	&	43.93 	&	0.303	&	44.56 	&	44.73 	&	44.02 	\\
J2347.0+5141	&	356.7659	&	51.6966	&	HBL	&	0.044	&	7.71 	&	42.85 	&	0.251	&	44.08 	&	43.67 	&	43.51 	\\
J2357.4-0152	&	359.3674	&	-1.8703	&	LBL	&	0.816	&	7.33 	&	43.91 	&	0.234	&	45.18 	&	45.91 	&	46.12 	\\
J2358.3+3830	&	359.5883	&	38.5097	&	IBL	&	0.2	&	7.39 	&	43.27 	&	0.0579	&	44.36 	&	44.30 	&	44.31 	\\
			\hline
            \end{tabular}
            \footnotesize{Columns (1) is the 4FGL name of sources; Columns (2) is the Right ascension in decimal degrees; Columns (3) is Declination in decimal degrees; Columns (4) is the Class of sources, low-frequency peaked BL Lac objects (LBL), intermediate objects (IBL), and high-frequency peaked BL Lac objects (HBL); Columns (5) is redshift; Columns (6) is the black hole mass; Columns (7) is the disk luminosity in units erg s$^{-1}$; Columns (8) is the 1.4 GHz radio flux in units jy; Columns (9) is the jet kinetic power in units erg s$^{-1}$; Columns (10) is the synchrotron peak frequency luminosity in units erg s$^{-1}$; Columns (11) is the inverse Compton luminosity in units erg s$^{-1}$. The redshift, black hole mass, accretion disk luminosity, synchrotron peak frequency luminosity and the inverse Compton luminosity comes from the work of \cite{Paliya2021}. The 1.4 GHz radio flux comes from the FIRST Survey Catalog: 14Dec17 Version\footnote{http://sundog.stsci.edu/first/catalogs/readme.html}.}
\end{minipage}
\end{table*}

\section{JET MODEL}
Currently, the most popular theories of jet formation include the BZ mechanism \citep{Blandford1977} and the BP mechanism \citep{Blandford1982}. Recently, some authors have proposed a hybrid jet model \citep{Meier1999, Garofalo2010}, that is, the mixture of BZ and BP. Theoretical works on jet power computation in the case of ADAF are often based on the self-similar solution of ADAFs of \cite{Narayan1995}, \cite{Meier2001}, and \cite{Nemmen2007}. There is some evidence to suggest that accretion flows on low accretion rate AGN (or low jet power, such as BL Lacs, low/hard-state X-ray binaries, and radio galaxies) are best described as ADAFs \citep[e.g.,][]{Wu2008, Wu2011}. In addition, the magnitude and structure of the magnetic field related to ADAFs are more favorable for extracting spin energy from the black hole than the magnetic field related to the standard thin disk \citep[e.g.,][]{Livio1999, Nemmen2007}. The BZ phenomenon is the backbone of the jet, and as the accretion increases, the BZ phenomenon becomes increasingly hybridized \citep{Cavaliere2002}. The jet mechanism becomes hybridized as the accretion rate increases. \cite{Tombesi2010} found that some relatively highly accreting radio galaxies have ultra-fast outflows with $v\sim(0.04-0.15)c$. Although these winds move at relativistic speeds, their findings indicate that some hydromagnetic windy activity is also present in AGNs with relativistic jets, thus enforcing the idea of a hybrid mechanism \citep{Foschini2011}. At the same time, our sample has a low accretion rate (see below), which implies that these sources may have ADAFs. Therefore, in this work, we calculate the jet power based on the self-similar solution of ADAFs around the Kerr black hole.         

\subsection{The BZ jet model}
The jet power of BZ mechanism ($P_{\rm jet}^{\rm BZ}$) is estimated by the following formula \citep{MacDonald1982, Thorne1986, Ghosh1997, Nemmen2007}

\begin{equation}
	P_{\rm jet}^{\rm BZ} = \frac{1}{32}\omega_{\rm F}^{2}B_{\perp}^{2}R_{\rm H}^{2}j^{2}c,
\end{equation}
where $\omega_{\rm F}\equiv \Omega_{\rm F}(\Omega_{\rm H}-\Omega_{\rm F})/\Omega_{\rm H}^{2}$ depends on the angular velocity of field lines $\Omega_{F}$ relative to that of the black hole $\Omega_{\rm H}$. We assume $\omega_{\rm F}=1/2$, which implies the output of maximum power \citep[e.g.,][]{MacDonald1982, Thorne1986, Nemmen2007}.
The $B_{\perp}$ is assumed to approximate to the poloidal component of the magnetic field $B_{\rm p}$, $B_{\perp}\approx B_{\rm p}(R_{\rm ms})\approx g(R_{\rm ms})B(R_{\rm ms})$\citep{Livio1999}, $g=\Omega/\Omega^{'}$. The $R_{\rm ms}$ is the radius of the marginally stable orbit of the accretion disk (see APPENDIX A). The $\Omega^{'}$ is the angular velocity of the disk. An observer at infinity will see the disk and the magnetic fields near the black hole rotate, in the Boyer–Lindquist coordinate system, not with an angular velocity $\Omega^{'}$ but $\Omega=\Omega^{'}+\omega$ \citep{Bardeen1972}, where $\omega$ is the angular velocity of the local metric which is given in Appendix A, equation (A11). The $B$ is the the magnetic field strength near the black hole (see APPENDIX A). The $R_{\rm H}=[1+(1-j^{2})^{1/2}]GM_{\rm BH}/c^{2}$ is the horizon radius \citep[e.g.,][]{Ghosh1997}, where $G$ is the gravitational constant and $M$ is the mass of the black hole. The $j$ is spin of black hole and $c$ is the speed of light.  

\subsection{The Hybrid jet model}
The jet power of the hybrid model ($P_{\rm jet}^{\rm Hybrid}$) is estimated by the following formula \citep{Meier2001, Nemmen2007}

\begin{equation}
	P_{\rm jet}^{\rm Hybrid} = (B_{\phi}HR\Omega)^{2}/32c, 
\end{equation}  
where $B_{\phi}$ is the azimuthal component of the magnetic field. \cite{Meier2001} related the amplified, azimuthal component of the magnetic field to the unamplified magnetic field strength derived from the self-similar ADAF solution as $B_{\phi}=gB$, and $\Omega=\Omega^{'}+\omega$. The $H$ is the vertical half-thickness of the disk. In the case of an ADAF, $H\sim R$ \citep{Nemmen2007}. Following the method of \cite{Nemmen2007}, all physical quantities are estimated at $R=R_{\rm ms}$ (see APPENDIX A).  

\section{Results and Discussion}
\subsection{Redshift and black hole masses}
To better evaluate our results and put them into a context, we studied the distributions of the Fermi BL Lacs of our sample in terms of redshift, black hole mass, jet kinetic power, and accretion disk luminosity. The redshift distribution of Fermi LBL extends to larger values than IBL and HBL (Fig.~\ref{figure0}). The average values of redshift are $z=0.64$ for LBL, $z=0.26$ for IBL, and $z=0.25$ for HBL. The range of black hole mass is from $10^{6.5}$ to $10^{10} M_{\odot}$. The average values of black hole mass are $\log M_{\rm BH}=8.39$ for LBL, $\log M_{\rm BH}=8.34$ for IBL, and $\log M_{\rm BH}=8.82$ for HBL. \cite{Cha2014} also found that FSRQs and LBL tend to have higher redshift and gamma-ray luminosity than HBL. They suggested that the evolutionary track of Fermi blazars is FSRQs$\rightarrow$LBL$\rightarrow$HBL.     

\subsection{Jet kinetic power and accretion disk luminosity}
\begin{figure}
	\includegraphics[width=8.5cm,height=8.5cm]{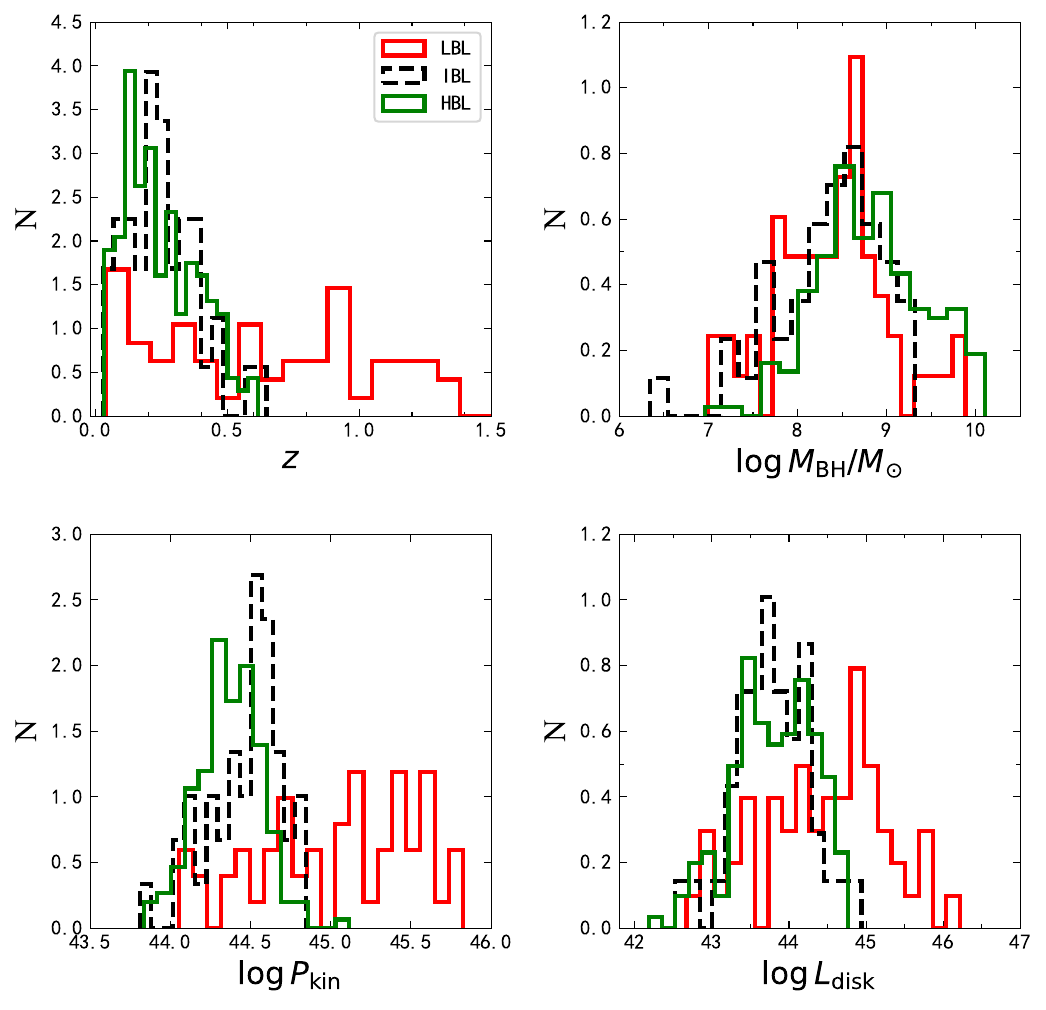}
	\caption{Distribution of the redshifts, black hole mass, jet kinetic power and accretion disk luminosity for the Fermi BL Lacs of our sample. The red line is LBL. The black dashed line is IBL. The green line is HBL.}
	\label{figure0}
\end{figure}

The range of jet kinetic power is mainly from $10^{44.0}$ to $10^{46.0}$ erg s$^{-1}$. The LBL tends to have higher jet kinetic power than IBL and HBL (Fig.~\ref{figure0}). The average values of jet kinetic power are $\log P_{\rm kin}=45.04$ for LBL, $\log P_{\rm kin}=44.26$ for IBL, and $\log P_{\rm kin}=44.37$ for HBL. The scope of accretion disk luminosity is mianly from $10^{42.0}$ to $10^{46.0}$ erg s$^{-1}$. The LBL also tends to have higher accretion disk luminosity than IBL and HBL. The average values of accretion disk luminosity are $\log L_{\rm disk}=44.45$ for LBL, $\log L_{\rm disk}=43.79$ for IBL, and $\log L_{\rm disk}=43.80$ for HBL. \cite{Chen2015a} found that the LBL tends to have higher jet kinetic power and BLR luminosity than HBL and IBL using a smaller sample than the one we consider here (see Figure 7 of \cite{Chen2015a}). Our results are consistent with theirs.

\subsection{Accretion rates}
\cite{Paliya2021} suggested that the physical properties of Fermi blazars are likely to be controlled by the accretion rate in Eddington units. Therefore, we study the accretion rate distribution of Fermi BL Lacs. Using the relative contribution of individual lines to the total BLR luminosity (see Section 2.2), we obtain the total line luminosity $L_{\rm lines}$. \cite{Wang2002} defined a "line accretion rate" and its dimensionless form as follows

\begin{eqnarray}
	\lambda = \frac{L_{\rm lines}}{L_{\rm Edd}}, L_{\rm lines}=\xi L_{\rm disk}.
\end{eqnarray}
Assuming that most of the line luminosity ($L_{\rm lines}$ ) is photoionized by the accretion disk \citep{Netzer1990}, the $L_{\rm lines}$ should be proportional to the total luminosity of the accretion disks. \cite{Netzer1990} defined $\xi\approx0.1$. The $L_{\rm Edd}$ is Eddington luminosity, $L_{\rm Edd}=1.3\times10^{38}(M_{\rm BH}/M_{\odot})\rm erg~s^{-1}$.  \cite{Wang2002} got the relation between $\lambda$ and the dimensionless accretion rate ($\dot{m}$) for an optically thin ADAFs as follows

\begin{equation}
	\dot{m}=2.17\times10^{-2}\alpha_{0.3}\xi_{-1}^{-1/2}\lambda_{-4}^{1/2},
\end{equation}
where $\alpha_{0.3}=\alpha/0.3$, $\xi_{-1}=\xi/0.1$, and $\lambda_{-4}=\lambda/10^{-4}$ \citep{Wang2003}. The viscosity parameters $\alpha$ is 0.3 \citep{Narayan1995}. \cite{Narayan1998} suggested that an optical thin ADAFs appears when $\dot{m}\leq\alpha^{2}$. Equation (8) can then be rewritten as

\begin{equation}
	\lambda_{1}=1.72\times10^{-3}\xi_{-1}\alpha_{0.3}^{2}.
\end{equation} 
Optically thin ADAFs require $\lambda<\lambda_{1}$. The optically thick, geometrically thin disk (SS) obey \citep{Wang2003}

\begin{equation}
	\dot{m}  = 10\xi_{-1}^{-1}\lambda,
\end{equation} 
and the condition $1>\dot{m}\geq\alpha^{2}$ gives

\begin{equation}
	\lambda_{2}=9.0\times10^{-3}\xi_{-1}\alpha_{0.3}^{2}.
\end{equation}
A standard thin disk satisfies $\lambda\geq\lambda_{2}$ \citep{Wang2003}. When the accretion rate reaches $\dot{m}\geq1$,  we have 

\begin{equation}
	\lambda_{3}=0.1\xi_{-1}.
\end{equation}   
A slim disk requires $\lambda\geq\lambda_{3}$ \citep{Wang2002, Wang2003}, namely Super-Eddington accretion (SEA). It is worth noting that in the transition region between $\lambda_{1}$ and $\lambda_{2}$, the accretion flow may be in a hybrid state, in which the standard disk coexists with ADAFs. Some authors have shown that hybrid states are possible in the accretion disk of AGN \citep[e.g.,][]{Quataert1999, Rozanska2000, Ho2000}. \cite{Gu2000} suggested that the conversion of a SS disk to an ADAFs is possible through evaporation \citep{Liu1999}. The transition radius depends on accretion rate, black hole mass, and viscosity. However, in such a regime, the disk structure is complex.   

\begin{figure}
	\includegraphics[width=8.5cm,height=8.5cm]{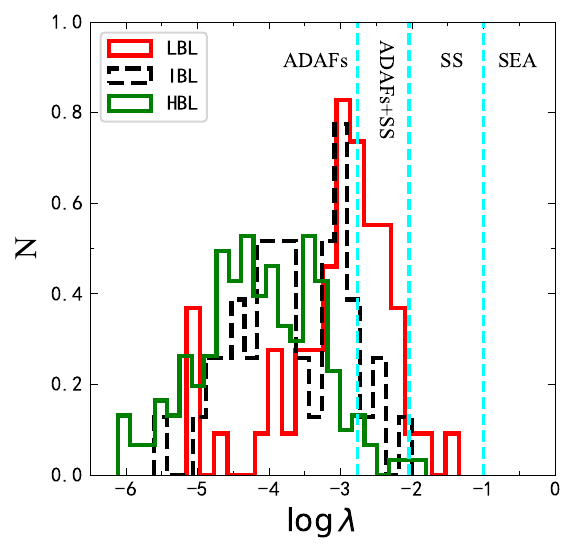}
	\caption{The distribution of $\lambda$ for the Fermi BL Lacs of our sample. Line styles identify a different class of BL Lacs as in Fig. 1. The cyan dashed lines are $\lambda_{1}$, $\lambda_{2}$, and $\lambda_{3}$, respectively. The distribution of $\lambda$ is divided into four regions, corresponding to different states of the accretion disks:(1) $\lambda<\lambda_{1}$ (pure ADAFs); (2) $\lambda_{1}\leq\lambda\leq\lambda_{2}$ (ADAFs+SS); (3) $\lambda_{2}\leq\lambda\leq\lambda_{3}$ (SS); (4) $\lambda\geq\lambda_{3}$ (SEA).}
	\label{figure3}
\end{figure}

Figure~\ref{figure3} shows that the three critical values of $\lambda$ define four regimes in accretion states. The average values of $\lambda$ for LBL, IBL and HBL are $\langle \log\lambda \rangle|_{\rm LBL} = -3.05$, $\langle \log\lambda \rangle|_{\rm IBL} = -3.67$ and $\langle \log\lambda \rangle|_{\rm HBL} = -4.14$, respectively. We find that the LBL has higher accretion rates than IBL and HBL, which could explain why LBL have more luminous disks and more powerful jets than HBL (see Fig.1). \cite{Wang2002} also found that the HBL has lower accretion rates than LBL. We investigated the differences in the distribution of accretion rates using the parameter T-test, nonparametric Kolmogorov Smirnov (K-S) test, and Kruskal Wallis H test. The parameter T-test is mainly used to test whether there is a difference in the mean accretion rate between two independent samples. The nonparametric K-S test and Kruskal Wallis H test are mainly used to test whether there are differences in the distribution of accretion rates between two independent samples. We assume that there are differences in the three tests at the same time, so there is a significant difference in the accretion rate between the two samples. According to the parameter T-test ($P = 1.11\times10^{-15}$, significant probability $P<0.05$), nonparametric K-S test ($P = 1.01\times10^{-15}$, significant probability $P<0.05$), and Kruskal–Wallis H-test ($P=1.99\times10^{-14}$, significant probability $P<0.05$), we find that the distributions of accretion rates between LBL and HBL are significantly different. The parameter T-test shows that there is a significant difference of the distributions of accretion rates between IBL and LBL ($P = 0.0002$). Through a nonparametric K-S test ($P = 0.0006$) and a Kruskal–Wallis H-test ($P = 9.38\times10^{-5}$), we find that the distributions of accretion rates between IBL and LBL are significantly different.  Through a parameter T-test ($P = 0.0008$), nonparametric K-S test ($P = 0.006$), and Kruskal–Wallis H-test ($P = 0.0008$), we find that the distributions of accretion rates between IBL and HBL are significantly different. 

From Figure~\ref{figure3}, we find that about 62\% HBL and 14\% IBL have pure optically thin ADAFs, while about 7\% LBL have ADAFs+SS. \cite{Cao2002} found that the accretion flows in all HBL are in the ADAF state. \cite{Wang2003} also found that the low-power HBLs and LBLs have pure optically thin ADAFs. However, the LBLs also may have a hybrid structure consisting of an SS disk plus optically thin ADAFs. \cite{Cao2003} also found that the BL Lacs have different accretion modes. They proposed that the BL Lacs have ADAFs in the inner region of the disk, and it becomes a standard thin disk in the outer region, i.e., ADAFs+SS scenario. 

\subsection{Relation between jet kinetic power and black hole mass}
\begin{figure}
	\includegraphics[width=8.5cm,height=8.5cm]{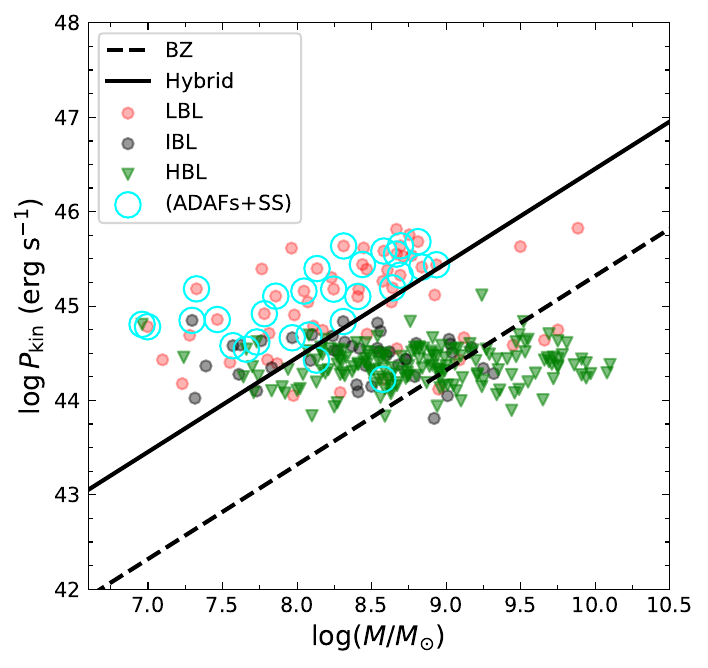}
	\caption{The jet kinetic power versus the black hole mass for the Fermi BL Lacs of our sample. The black dashed line is the maximum jet power expected in the case of BZ jet formation mechanism. The black solid line is the maximum jet power expected in the case of Hybrid jet formation mechanism. The red dot is LBL. The black dot is IBL. The green trangle is HBL. The cyan circle is sources with ADAFs+SS.}
	\label{figure1}
\end{figure}

Because our sample has an accretion rate compatible with an ADAFs regime (Fig~\ref{figure3}), we considered the jet power of the BZ mechanism and the jet power of the hybrid model in the ADAFs scenario. The relation between the jet power extracted from the ADAFs for the BZ and hybrid models and the black hole mass is shown in Figure~\ref{figure1}. In order to estimate the maximum jet power of BZ model and Hybrid model, we use accretion rates $\dot{m}=0.01$, and the disk viscosity parameter $\alpha=0.3$ \citep{Narayan1995}. The spin of black hole $j=0.98$ is adopted. The black dashed line is the BZ jet model, and the black solid line is the Hybrid jet model. The red dot is LBL. The black dot is the IBL. The green triangle is HBL. The cyan circle is sources with ADAFs+SS. We find that about 7\% LBL, 9\% IBL, and 32\% HBL are below the maximal BZ jet power expected to be extracted from ADAFs (Fig.\ref{figure1}, dashed line). These results show that the jet kinetic power of LBL and IBL can hardly be explained by the BZ jet model. However, about 26\% LBLs, 72\% IBLs, and 94\% HBLs are below the solid line when we consider the hybrid model. These results show that most of the IBL and HBL can be explained by the Hybrid jet model. Most LBLs cannot be explained by the BZ jet model or Hybrid jet model. These sources with ADAFs+SS have high jet power, which makes them unable to be explained by BZ or hybrid models. There are two possible explanations for the jet power of the LBL. One is that these LBLs require other jet models, such as an accretion disk with magnetization-driven outflows \citep{Cao2013, Li2014, Cao2016, Cao2018, Li2022}. \cite{Cao2013} proposed that if the most angular momentum of the gas in the thin disk is taken away by the magnetically driven outflows, the radial velocity of the disk will increase significantly, so the external field can be significantly enhanced in the inner region of the thin disk with the magnetically driven outflows \citep{Li2019, Cao2018}. Thereby, the expected jet kinetic power is higher than in the case of the BZ or hybrid model, as we observed in the data. The other is that the LBL has a strong beaming effect. \cite{Lister2011} found that the LBL has generally higher Doppler factors than HBL, which implies that the LBL has a strong beaming effect. Because of the beaming effect, the jet kinetic power could be overestimated. The "real" jet power, in this scenario, may be lower and compatible with models.

It is widely believed that large-scale magnetic fields play a crucial role in the acceleration and collimation of jets and/or outflows \citep[e.g.,][]{Pudritz2007}. The origin of the large-scale magnetic field passing through the accretion disk has not been well understood. Some studies suggest that large-scale magnetic fields that accelerate jets or outflows may be formed by weak external field advection \citep[e.g.,][]{Bisnovatyi1977, Bisnovatyi1976, Spruit2005}. However, \cite{Lubow1994} found that in the geometrically thin accretion disk (H/R $\leq$ 1), the advection in the external field is quite ineffective because of its small radial velocity. This may imply that the field in the inner region of the disk is not much stronger than the external weak field, which cannot accelerate strong jets in radio-loud quasars \citep{Lubow1994}. A few mechanisms were suggested to alleviate the difficulty of field advection in the thin disk \citep[e.g.,][]{Spruit2005}. Some people believe that the hot corona above the disk can effectively drag the external field inward, that is, the so-called "coronal mechanism" \citep[see][]{Beckwith2009}. The radial velocity of the gas above the disk can be greater than the radial velocity of the mid-plane of the disk, which partly solves the problem of inefficient field advection in the thin disk \citep{Lovelace2009, Guilet2012, Guilet2013}. Recently, \cite{Cao2018} suggested that powerful jets can be accelerated by the coronal magnetic field. \cite{Cao2004} found that the jets are accelerated from the disk coronas for radio-loud quasars. \cite{Zhu2020} found a a close connection between corona-disk-jet for radio-loud quasars. At the same time, some authors found that FSRQs and LBL have similar spectral properties, such as particle acceleration mechanism \citep[e.g.,][]{Chen2021}. These results indicate that the jet kinetic power of LBL may also be explained by the coronal magnetic field. 

\cite{Chen2023a} found that the jet kinetic power of jetted AGNs are powered by the BZ mechanism based on the relation between jet kinetic power and accretion disk luminosity. Our results are slightly different from those of \cite{Chen2023a}. The main reason is that we calculate the maximum jet power of the BZ mechanism and the hybrid model based on the theory, and then compare the maximum jet power of the BZ mechanism and the hybrid model with the observed jet kinetic power. The work of \cite{Chen2023a} is mainly based on whether the slope of the relationship between jet kinetic power and accretion disk luminosity is 1 to judge whether their jet knetic power is dominated by the BZ mechanism. In the work of \cite{Chen2023a}, it is assumed that the jet kinetic power of jetted AGN is dominated by Poynting flux, and then it is obtained that the slope of the relationship between the maximum jet power of BZ mechanism and the luminosity of accretion disk is equal to 1 \citep{Ghisellini2006}. Finally, the slope of the relationship between jet kinetic power and accretion disk luminosity is compared the slope obtained with theory to further judge whether the jet of jetted AGN is dominated by BZ mechanism. However, some studies have found that the jet power of jetted AGN is not dominated by the Poynting flux \citep[e.g.,][]{McKinney2012, Zdziarski2015, Paliya2017, Chen2023b}.                    

\begin{figure}
	\includegraphics[width=8.5cm,height=8.5cm]{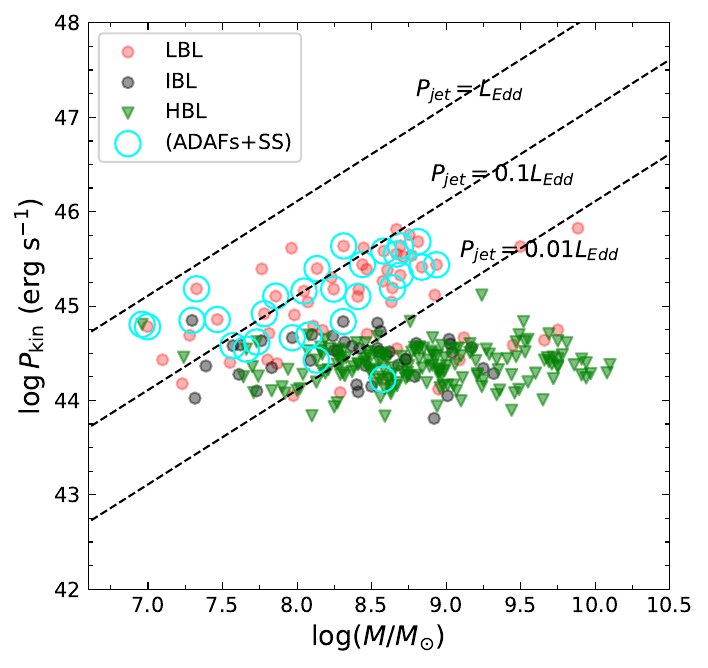}
	\caption{The jet kinetic power versus the black hole mass. The dashed lines are
		$P_{\rm jet}/L_{\rm Edd}$ = 0.01, 0.1, 1, respectively. The different symbols are as in Fig.3.}
	\label{figure2}
\end{figure}

In Figure~\ref{figure2}, we show the results as in Fig~\ref{figure1}, but the dashed lines are $P_{\rm jet}/L_{\rm Edd}$ = 0.01, 0.1, 1, respectively. The jet kinetic power of all Fermi BLLacs is less than 1$L_{\rm Edd}$. We find that about 79\% LBL is located in $\sim0.01-1L_{\rm Edd}$. As a comparison, the jet power expected in case of coronal mechanism is at most 0.05 Eddington \citep{Cao2018}. \cite{Cao2013} suggested that most of the angular momentum of the accretion disk is excluded from magnetization-driven outflows, and the magnetic field will be enhanced compared with the thin disk without outflow. The magnetic field dragged inward by the accretion disc with magnetization-driven outflows may accelerate the jet in source with high jet power. The jet kinetic power of these LBLs may be explained by the magnetization-driven outflows model \citep{Cao2018}.     

We also find that the jet kinetic power of LBL depends on the mass of the black hole, while the jet kinetic power of HBL does not seem to depend on the mass of the black hole. \cite{Ghosh1997} found that it is possible for the standard disks to find two regimes, which may be radiation pressure dominated (RPD) or gas pressure dominated (GPD). The jet kinetic power depends on the mass of the black hole, indicating that the accretion disk is dominated by radiation pressure. The jet kinetic power depends on the accretion ($L_{\rm bol}/L_{\rm Edd}$), indicating that the accretion disk is dominated by gas pressure \citep{Ghosh1997, Foschini2011, Chen2015a}. Our results imply that the accretion disk of LBL may be dominated by radiation pressure, while that of HBL is not dominated by radiation pressure. The accretion disks of HBL may be dominated by gas pressure. In the future, when our source has the bolometric luminosity, we will test these results.  \cite{Chen2015a} also found that the accretion disks of FSRQs and LBL are dominated by radiation pressure, while the accretion disks of HBL and IBL seem to be dominated by gas pressure.    

\subsection{Relation between jet kinetic power and accretion disk luminosity} 
\begin{figure}
	\includegraphics[width=8.5cm,height=8.5cm]{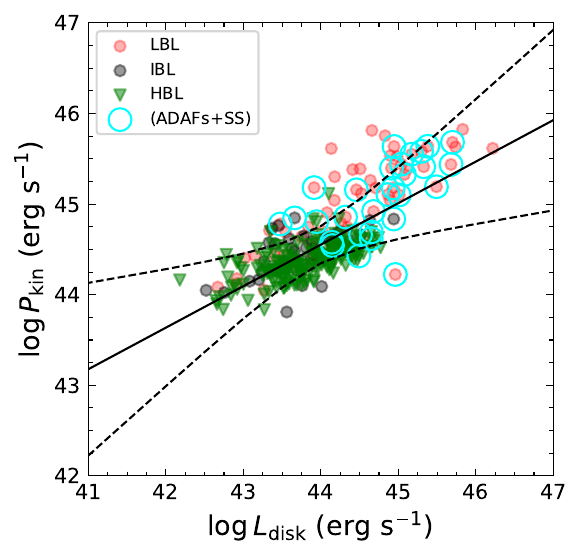}
	\caption{Relation between jet kinetic power and accretion disk luminosity for the Fermi BL Lacs of our sample. The different symbols are as in Fig.3. The solid line corresponds to the best-fitting linear models. The dashed lines indicate 3$\sigma$ confidence bands of the best fits, namely it refers to the significance level $a=0.01$, with a 99\% probability that the measured results are within this range.}
	\label{Figure4}
\end{figure}

There is evidence that there is a positive correlation between jet power and accretion luminosity in jetted AGN \citep[e.g.,][]{Rawlings1991, Wang2004, Gu2009, Ghisellini2010, Sbarrato2014, Ghisellini2014, Paliya2019}. The relation between jet kinetic power and accretion disk luminosity is shown in Figure~\ref{Figure4}. We find a significant correlation between jet kinetic power and accretion disk luminosity for the whole sample ($r=0.74, P=5.10\times10^{-50}$). The tests of Spearman ($r = 0.69, P = 1.58\times10^{-41}$) and Kendall tau ($r = 0.52, P = 3.81\times10^{-37}$) also show a significant correlation between jet kinetic power and accretion disk luminosity for the whole sample. The best fitting equation given by least square linear regression is $\log P_{\rm kin}=(0.46\pm0.03)\log L_{\rm disk}+(24.39\pm1.09)$. At the same time, we also find that the sources with ADAFs+SS follow the same relation. However, \cite{Rajguru2022} suggested that the correlations are often driven by the common redshift dependence. They found a weak correlation between jet power and accretion disk luminosity when the redshift was excluded. We test their results. Partial correlation shows a moderately weak correlation between jet kinetic power and accretion disk luminosity when redshift dependence is excluded for the whole sample ($r=0.12, P=0.03$). We use Pearson correlation analysis for other types of AGNs. There is {\bf also} a significant correlation between jet kinetic power and accretion disk luminosity for the LBL ($r=0.81, P=3.51\times10^{-14}$). The best fitting equation given by least square linear regression for the LBL is $\log P_{\rm kin}=(0.48\pm0.05)\log L_{\rm disk}+(23.46\pm2.13)$. Partial correlation shows a moderately weak correlation between jet kinetic power and accretion disk luminosity when redshift dependence is excluded for the LBL ($r=0.34, P=0.009$).  There is a significant correlation between jet kinetic power and accretion disk luminosity for the IBL+HBL ($r=0.60, P=3.09\times10^{-23}$). The best fitting equation given by least square linear regression for the IBL+HBL is $\log P_{\rm kin}=(0.25\pm0.02)\log L_{\rm disk}+(33.42\pm0.98)$. Partial correlation shows a moderately weak correlation between jet kinetic power and accretion disk luminosity when redshift dependence is excluded for the IBL+HBL ($r=0.24, P=0.003$). We find that the slope of the relation between jet kinetic power and accretion disk luminosity for the LBL is slightly different from that of IBL+HBL, and the slope of LBL is greater than that of IBL+HBL. \cite{Paliya2017} found that the slope of the relation bteween jet power and accretion disk luminosity is 0.46 using both 324 $\gamma$-ray detected and 191 $\gamma$-ray undetected blazars. \cite{Chen2023b} also found that the slope of the relation bteween jet kinetic power and accretion disk luminosity is $0.51\pm0.18$ using 38 gamma-ray-emitting radio galaxies. Our results are similar to the results of \cite{Paliya2017} and \cite{Chen2023b}. \cite{Ghisellini2014} found that the slope of the relation bteween jet power and accretion disk luminosity is 0.92 using 217 $\gamma$-ray detected blazars. \cite{Chen2023a} found that the slope of the relation bteween jet kinetic power and accretion disk luminosity is $1.00\pm0.02$ for the whole sample (FSRQs+BL Lacs+gamma-ray-emitting narrow-line Seyfert 1 galaxies),  $0.83\pm0.04$ for FSRQs,  $1.00\pm0.05$ for BL Lacs, $0.73\pm0.15$ for gamma-ray-emitting narrow-line Seyfert 1 galaxies. Because we mainly use the jet kinetic power to replace the total jet power. Therefore, we also examined all relationships including 10\% of jet kinetic power and found that all relationships were valid in our work.        

\subsection{Relation between IC Luminosity versus Synchrotron Luminosity}
\begin{figure}
	\includegraphics[width=8.0cm,height=8.0cm]{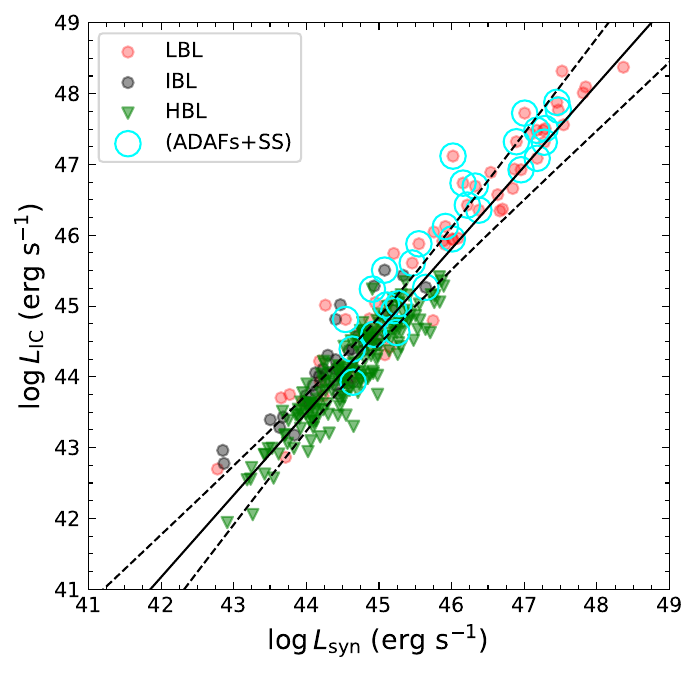}
	\caption{The inverse Compton luminosity versus synchrotron luminosity for the Fermi BL Lacs of our sample. The different symbols are as in Fig.3. The solid and dashed lines describe the best fit linear model and its uncertainty as in Fig. 5.}
	\label{Figure5}
\end{figure}

Figure~\ref{Figure5} shows a relation between IC luminosity and synchrotron
luminosity for Fermi BL Lacs. We find a significant correlation between IC luminosity and synchrotron luminosity for the whole sample ($r=0.95, P=2.06\times10^{-143}$). The tests of Spearman ($r = 0.92, P = 7.66\times10^{-113}$) and Kendall tau ($r = 0.76, P = 3.47\times10^{-80}$) also show a significant correlation between IC luminosity and synchrotron luminosity for the whole sample. The best fitting equation given by least square linear regression is $\log L_{\rm IC} = (1.16\pm0.02)\log L_{\rm syn} + (-7.58\pm1.00)$. We find that the slope of the relation between IC luminosity and synchrotron luminosity is close to 1.0. According to \cite{Ghisellini1996}($L_{\rm EC}\sim L_{\rm syn}^{1.5}$, $L_{\rm SSC}\sim L_{\rm syn}^{1.0}$), our results suggest that the IC component of Fermi BL Lacs is dominated by the SSC process. Some authors have confirmed this conclusion \citep[e.g.,][]{Celotti2008, Ghisellini2010, Lister2011, Ackermann2012, Wu2014, Marchesini2019, LaMura2022}. \cite{Xue2016} found that the slope of the relation between IC luminosity and synchrotron luminosity for 28 Fermi BL Lacs is $k_{\rm BL Lacs}=1.12\pm0.10$. Our work confirms the results of previous studies.     

\section{SUMMARY}
We use a large sample of Fermi BL Lacs to study the physical properties of their jets, and the main conclusions are as follows:

(1) We find that that LBL tends to have higher accretion disk luminosity and jet kinetic power than HBL, which could be due to a higher accretion rate.

(2) We find that LBL has a higher accretion rate than IBL and HBL. Almost all IBL and HBL have pure optically thin ADAFs. However, some LBLs may have a hybrid structure consisting of an SS disk plus optically thin ADAFs. 

(3) We study the jet power of the BZ mechanism and the hybrid mechanism based on the self-similar solution of the ADAFs around the Kerr black hole. Through the relationship between jet power and black hole mass, we find that the jet kinetic power of about 72\% IBL and 94\% HBL can be explained by the hybrid model. However, only 7\% LBL can be explained by the BZ model, and only 26\% LBL can be explained by the hybrid model. 

(4) The jet kinetic power of about 79\% LBL is located at $\sim0.01-1L_{\rm Edd}$. The jet kinetic power of these sources may be interpreted by magnetization-driven outflows. 

(5) There is a significant correlation between jet kinetic power and accretion disk luminosity for Fermi BL Lacs. However, this correlation becomes weak when excluding the dependence of redshift. 

(6) We find a significant correlation between IC luminosity and synchrotron luminosity for Fermi BL Lacs. The slope of the relation between IC luminosity and synchrotron luminosity in Fermi BL Lacs is $1.16\pm0.02$, which implies that the high-energy components of Fermi BL Lacs are dominated by the SSC process.
 
\section*{Acknowledgements}
We are very grateful to the referee and Editor for the very helpful report. Yongyun Chen is grateful for financial support from the National Natural Science Foundation of China (No. 12203028). This work was support from the research project of Qujing Normal University (2105098001/094). This work is supported by the youth project of Yunnan Provincial Science and Technology Department (202101AU070146, 2103010006). Yongyun Chen is grateful for funding for the training Program for talents in Xingdian, Yunnan Province. 
QSGU is supported by  the National Natural Science Foundation of China (No. 12192222, 12192220 and 12121003).
We also acknowledge the science research grants from the China Manned Space Project with NO. CMS-CSST-2021-A05. This work is supported by the National Natural Science Foundation of China (11733001 and U2031201). D.R.X. acknowledge the science research grants from the China Manned Space Project with No. CMS-CSST- 2021-A06, Yunnan Province Youth Top Talent Project (YNWR-QNBJ-2020-116) and the CAS “Light of West China” Program.

\section*{Data Availability}
All the data used here are available upon reasonable request. All datas are in Table 1. 

\bibliographystyle{mnras}
\bibliography{example} 




\appendix

\section{Derivation of the jet power}
\label{appenda}
We list all the equations we use to calculate the jet power depending on $\alpha$ (viscosity parameter), $j$ (spin of black hole) and $\dot{M}$ (accretion rate on to the black hole) using the BZ model (Section3.1) and the hybrid model (Section 3.2). To estimate the maximum jet power, $\alpha=0.3$ and $j=0.98$ are adopted. The jet power is given by equation (7) (BZ model) and equation (8) (hybrid model). The self-similar ADAFs structure is described by \cite{Narayan1995}.  We use the black hole mass in solar units ($m=M_\bullet/M_\odot$), accretion rates in Eddington units ($\dot{m}=\dot{M}/\dot{M}_{\rm Edd}$, $\dot{M}_{\rm Edd}$ is the Eddington accretion rate ($\dot{M_{\rm Edd}}\equiv22M_{\bullet}/(10^{9}M_{\odot})$), $\dot{m}=0.01$ is adopted \citep{Narayan1995}) and radii in Schwarzschild units ($r=R/(2 GM_\bullet/c^2 )$):

\begin{eqnarray}
	\Omega' = 7.19 \times 10^4 c_2 m^{-1} r^{-3/2} \; {\rm s}^{-1}, \\
	B = 6.55 \times 10^8 \alpha^{-1/2} (1-\beta)^{1/2} c_1^{-1/2} c_3^{1/4} m^{-1/2} \dot{m}^{1/2} r^{-5/4} \; {\rm G}, \\
	H/R \approx (2.5 c_3)^{1/2}.
\end{eqnarray} 
The $\Omega^{'}$ is  the angular velocity of the disk, $B$ is the magnetic field strength near the black hole in terms of radius R, $G$ is the gravitational constant, and $H$ is the vertical half-thickness of the disk. The constants $c_1$, $c_2$ and $c_3$ are defined as
\begin{eqnarray}
	c_1 = \frac{5+2\epsilon'}{3 \alpha^2} g'(\alpha,\epsilon'), \\
	c_2 = \left[  \frac{2 \epsilon' (5+2\epsilon')}{9 \alpha^2} g'(\alpha,\epsilon') \right] ^{1/2}, \\
	c_3 = c_2^2/\epsilon', \\
	\epsilon' \equiv \frac{1}{f} \left( \frac{5/3-\gamma}{\gamma-1} \right), \\
	g'(\alpha,\epsilon') \equiv \left[ 1+ \frac{18 \alpha^2}{(5+2\epsilon')^2} \right]^{1/2}. 
\end{eqnarray} 
where advection parameter $f$ (assumed $\approx$ 1, see \cite{Nemmen2007}), $\epsilon'$ and $g'(\alpha,\epsilon')$ are just a variable replacement. The $\gamma$ is adiabatic index of the accreting fluid. 

The relationship between $\alpha$, $\beta$ and $\gamma$ is defined as follows

\begin{eqnarray}
	\gamma=(5\beta+8)/3(2+\beta), \\
	\alpha\approx0.55/(1+\beta).
\end{eqnarray}
where $\alpha$ is viscosity parameter and $\beta$ is the ratio of gas to magnetic pressure \citep{Esin1997, Hawley1995}. The angular velocity of the field seen by an outside observer at infinity in the Boyer-Lindquist frame is $\Omega = \Omega' + \omega$. \cite{Bardeen1972} gave the formula for calculating the angular velocity of the local metric as follows  

\begin{equation}
	\omega=\frac{2jM_{\bullet}}{j^{2}(R+2M_{\bullet})+R^{3}}
\end{equation}
using geometrized units ($G = c = 1$). \cite{Meier2001} estimated the field-enhancing shear caused by the Kerr metric using $g=\Omega/\Omega'$, such that the azimuthal component of the field is given by $B_\phi = g B$. The poloidal component is related to the azimuthal component following \cite{Livio1999} as $B_p \approx H/R \; B_\phi \approx B_\phi$. In the case of an ADAFs,  this yields $B_p\approx B_\phi$ because $H\sim R$. The advection parameter $f$ is assumed $f\approx$1 \citep{Nemmen2007}. We insert all the quantities defined above into Equations (7) and (8), and then evaluate the resulting equations at the marginally stable orbit of the accretion disk $R_{\rm ms}$. \cite{Bardeen1972} defined the marginally stable orbit of the accretion disk ($R_{\rm ms}$) as follows 
\begin{eqnarray}
	R_{\rm ms} =GM_{\bullet}/c^{2}\{3+Z_{2}-\left[(3-Z_{1})(3+Z_{1}+2Z_{2})\right]^{1/2}\}, \\
	Z_{1}\equiv1+(1-j^{2})^{1/3}\left[(1+j)^{1/3}+(1-j)^{1/3}\right], \\
	Z_{2}\equiv(3j^{2}+Z_{1}^{2})^{1/2}.
\end{eqnarray}


\bsp	
\label{lastpage}
\end{document}